\documentclass[apj]{emulateapj}
\usepackage{apjfonts}
\journalinfo{The Astrophysical Journal, in press}
\submitted{Received 2005 September 22; accepted 2005 November 18}

\shorttitle{Global properties of Draco}
\shortauthors{Mashchenko et al.}

\begin{document}

\title{Constraining global properties of the Draco dwarf spheroidal galaxy}

\author{Sergey Mashchenko, Alison Sills, and H. M. P. Couchman}

\affil{Department of Physics and Astronomy, McMaster University,
Hamilton, ON, L8S 4M1, Canada; syam,couchman,asills@physics.mcmaster.ca}

\begin{abstract}
By fitting a flexible stellar anisotropy model to the observed surface brightness and
line-of-sight velocity dispersion profiles of Draco we derive a sequence of
cosmologically plausible two-component (stars $+$ dark matter) models for
this galaxy. The models are consistent with all the available observations and can
have either cuspy Navarro-Frenk-White or flat-cored dark matter density profiles.
The dark matter halos either formed relatively recently (at $z\sim 2\dots 7$)
and are massive (up to $\sim 5\times 10^9$~M$_\odot$), or formed before the end of
the reionization of the universe ($z\sim 7\dots 11$) and are less massive (down
to $\sim 7\times 10^7$~M$_\odot$). Our results thus support either of the two
popular solutions of the ``missing satellites'' problem of $\Lambda$ cold dark matter
cosmology -- that dwarf spheroidals are either very massive, or very old.  We carry out
high-resolution simulations of the tidal evolution of our two-component Draco
models in the potential of the Milky Way. The results of our simulations suggest
that the observable properties of Draco have not been appreciably affected by
the Galactic tides after 10~Gyr of evolution.  We rule out Draco being a ``tidal
dwarf'' -- a tidally disrupted dwarf galaxy.  Almost radial Draco orbits (with
the pericentric distance $\lesssim 15$~kpc) are also ruled out by our analysis.
The case of a harmonic dark matter core can be consistent with observations only for a
very limited choice of Draco orbits (with the apocentric-to-pericentric
distances ratio of $\lesssim 2.5$).

\end{abstract}


\section{INTRODUCTION}

Galactic dwarf spheroidal galaxies (dSphs) are intriguing objects with
a deceptively simple appearance -- roughly spheroidal shape, no gas, and no recent
star formation.  Due to their closeness, these galaxies are studied in
significantly more detail than other external galaxies (see the review of
\citealt{mat98}).  Despite that, the nature of dSphs and their place in the
larger cosmological picture is not well understood.

The first wave of enhanced interest in dSphs took off after the pioneer work of
\citet{aar83}. Based only on the three stars in Draco with measured
line-of-sight velocities, he boldly claimed that Draco can be significantly dark
matter (DM) dominated. The fact that the stellar velocity dispersion in dSphs is
significantly larger than what would follow from the virial theorem for the
luminous mass was later confirmed with much larger studies.  For example, the
latest compilation of \citet{wil04} contained 207 Draco stars with measured
line-of-sight velocities. Similar results were obtained for other dSphs as well
-- both for the Milky Way and M31 satellites. There were attempts to explain the
large stellar velocity dispersion in Galactic dSphs without invoking the DM
hypothesis. Most notably, \citet{kro97} presented a model where dSphs are
considered to be ``tidal dwarfs'' -- virtually unbound stellar streams from
dwarf galaxies tidally disrupted in the Milky Way potential. The model of
\citet{kro97} appears to be not applicable to Draco as it is unable to reproduce
the narrow observed horizontal branch width of this dwarf \citep{kle03}. In this
paper we will present additional evidence against Draco being a tidal dwarf. The
overall consensus now appears to be that dSphs do contain significant amounts of
DM and are hence the smallest known objects with (indirectly) detected DM.  This
makes them very interesting objects, as they can be an important test bench for
modern cosmological models and for the theories of DM.

More recently, the interest in dSphs was rejuvenated after the realization that
simple (DM-only) cosmological models overpredict the number of the Milky Way
satellite galaxies by $1-2$ orders of magnitude \citep{kly99,moo99}.  This was
coined the ``missing satellite problem'' of cosmology.  The original analysis
was done under the assumption that DM in dSphs has the same spatial extent as
stars.  Relaxation of the above assumption led to a suggestion that only the
most massive subhalos predicted to populate a Galaxy-sized halo managed to form
stars, with the rest of the subhalos staying dark \citep{sto02}. Another way of
solving the missing satellites problem is to assume that only the oldest
subhalos formed stars, with the star formation in the younger subhalos being
suppressed by the metagalactic ionizing radiation after the reionization of the
universe was accomplished around $z\sim 6.5$ \citep{bul00}. In reality, 
both mechanisms could have realized \citep{ric05}.

To discriminate between the two above solutions of the missing satellites
problem and to place Galactic dSphs in the right cosmological context one has to
know the global parameters of these dwarfs, and most importantly, their total DM
extent and mass. The traditional approach is to assume that the dwarf is in
equilibrium (thus ignoring the possible impact of the Galactic tides), and to
solve the Jeans equation to infer the density profile of the DM halo based on
the observed surface brightness and line-of-sight velocity dispersion
profiles. As the proper motions of individual stars in dSphs are not known, one
has to resort to making certain assumptions about the anisotropy in the stellar
velocities.  Due to a well known degeneracy between the assumed stellar
anisotropy and inferred total enclosed mass there are many solutions to the
Jeans equation which are consistent with the observations. Another limitation of
the above approach is that DM can be traced only within the stellar body extent
of the dwarf, so no conclusion can be made about the total mass of the
galaxy. It is also not clear at what distance from the dwarf's center the virial
equilibrium assumption breaks down due to Galactic tides. Some work has been
done on the impact of tidal forces on the structure of Galactic satellites
\citep*[e.g.][]{oh95,pia95,hay03,kaz04}, where it was clearly demonstrated
that the tidal stripping and shocking is a complex dynamic process.  The
deficiency of the above work is in using single-component models for the dwarf
galaxies, which made it impossible to directly compare the results of the
numerical simulations with the observations. In general, stars in dSphs
are distributed differently from DM, so they also behave differently in reaction to
the external tides. To correctly describe the observable manifestations of the
Galactic tides in dSphs it is essential to use two-component (stars $+$ DM)
dwarf models \citep{rea05}.

In this paper we place joint constraints on the global properties of Draco (one
of the best studied dSphs) by (1) using cosmological predictions for the
properties of DM halos, (2) developing very flexible stellar anisotropy model
for dSphs, and (3) running a large set of high-resolution simulations of the
evolution of two-component dwarfs in the Galactic tidal field. We derive a
sequence of cosmologically plausible models for Draco which are consistent with
all the observed structural and kinematical properties of this dwarf.  Our
results are consistent with either of the two above solutions for the missing
satellite problem.

\section{Global Constraints on Draco DM Halo Properties}
\label{global}

We consider two types of DM halos: theoretically motivated \citet*[hereafter NFW]{NFW97} halos
with a $\gamma=-1$ density cusp at the center,
and observationally motivated \citet{bur95} halos,
which have a flat core:

\begin{eqnarray}
\rho(r)& =& \frac{\rho_s}{r/r_s\, (1+r/r_s)^2}\quad\mbox{(NFW),}\label{eq1}\\
\rho(r)& = &\frac{\rho_s}{(1+r/r_s)\,[1+(r/r_s)^2]}\quad\mbox{(Burkert).}\label{eq2}
\end{eqnarray}

\noindent Here $\rho_s$ and $r_s$ are the scaling density and radius.
At large distances from the center both halos have the same asymptotic density
slope of $\gamma=-3$.

Analysis of cosmological $N$-body simulations showed that the concentration
$c=r_{\rm vir}/r_s$ of low-mass DM halos (with the virial mass $m_{\rm vir}=10^8\dots
10^{11}$~M$_\odot$) has a log-normal distribution with the mean

\begin{equation}
\label{eqc}
c=\frac{27}{1+z} \left(\frac{m_{\rm vir}}{10^9 M_\odot}\right)^{-0.08}
\end{equation}

\noindent and dispersion 0.14~dex \citep[with the correction of \citealt*{ste02}]{bul01}.
Here $r_{\rm vir}$ is the virial radius of the halo and $z$ is the redshift.

\citet{ste02} showed that the four dwarf galaxies with a Burkert DM halo profile studied by \citet{bur95}  have
the same dependence of the DM halo scaling density $\rho_s$ on the scaling
radius $r_s$ as do the NFW halos in cosmological simulations. This result was
obtained for $z=0$.  We assume that it holds true for other redshifts as well,
and use the equation~(\ref{eqc}) to find concentrations of both NFW and Burkert
halos.

Using the formula of \citet{she99}, one can estimate the comoving number density of DM halos
per unit $\ln m_{\rm vir}$ and per standard deviation in concentration:

\begin{eqnarray}
\label{ST2}
F\equiv \frac{dn}{d\ln m_{\rm vir} d\nu_c}&&\nonumber\\
= \frac{0.322}{2\pi} \left(1+\frac{1}{\nu^{0.6}}\right) \frac{\rho_m \nu}{S}\left|\frac{dS}{dm_{\rm vir}}\right|
\exp\left(-\frac{\nu^2+\nu_c^2}{2}\right).&&
\end{eqnarray}

\noindent Here $\nu_c$ is the number of standard deviations from the mean concentration,
$\nu= (0.707/S)^{1/2} \delta(z)$, where $\delta(z)$ is the critical overdensity
for spherical collapse extrapolated linearly to $z=0$, $\rho_m$ is the present
day mass density of the universe, and $S$ is the variance of the primordial
density field on mass scale $m_{\rm vir}$ extrapolated linearly to $z=0$. To
estimate the above parameters we follow \citet*{mcs05}. Throughout this paper we
assume a flat $\Lambda$CDM cosmology and use the following values of the
cosmological parameters: $\Omega_m=0.27$, $\Omega_b=0.044$,
$H=71$~km~s$^{-1}$~Mpc$^{-1}$, and $\sigma_8=0.84$
\citep{spe03}.

\begin{figure*}
\plottwo{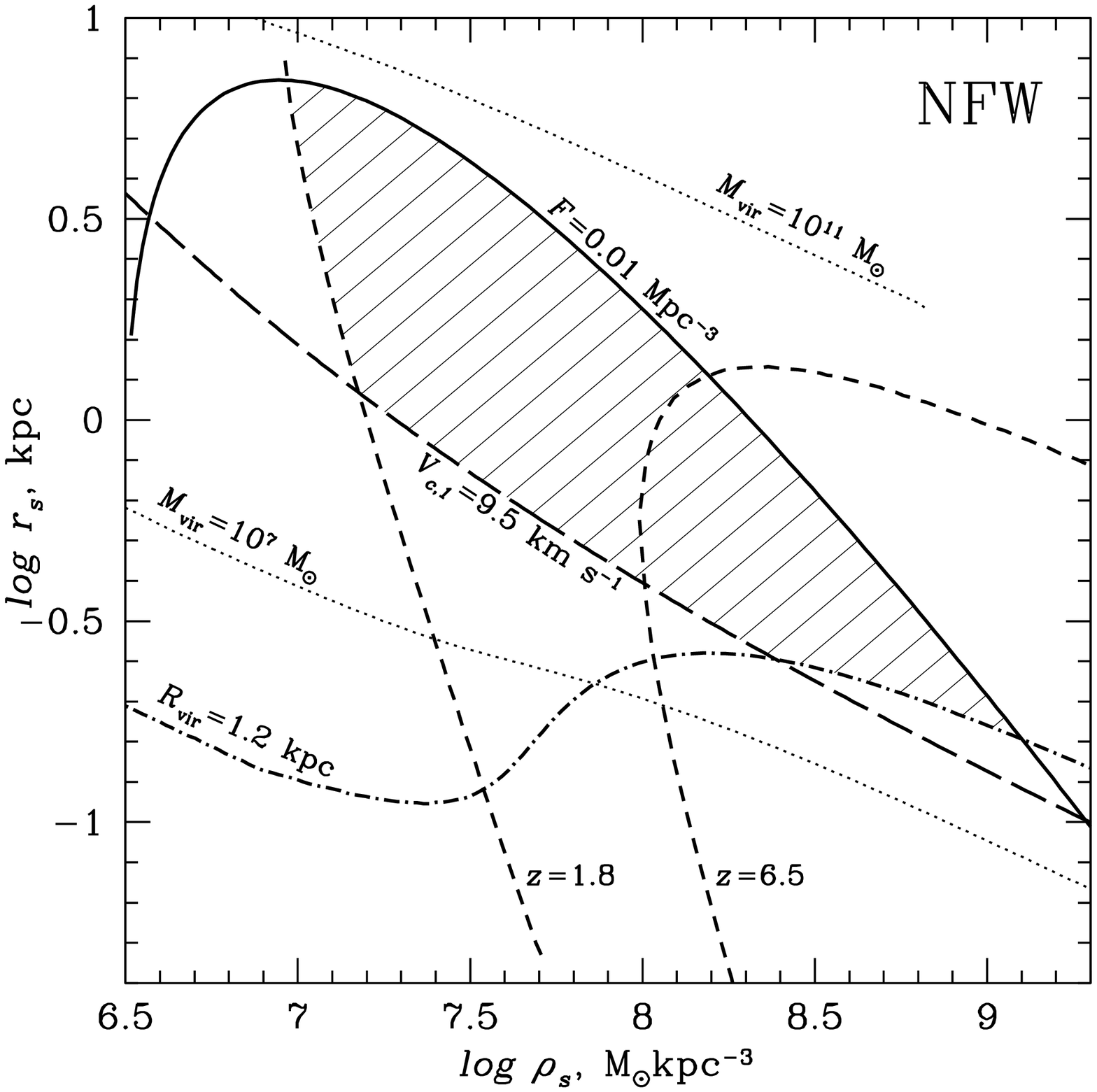}{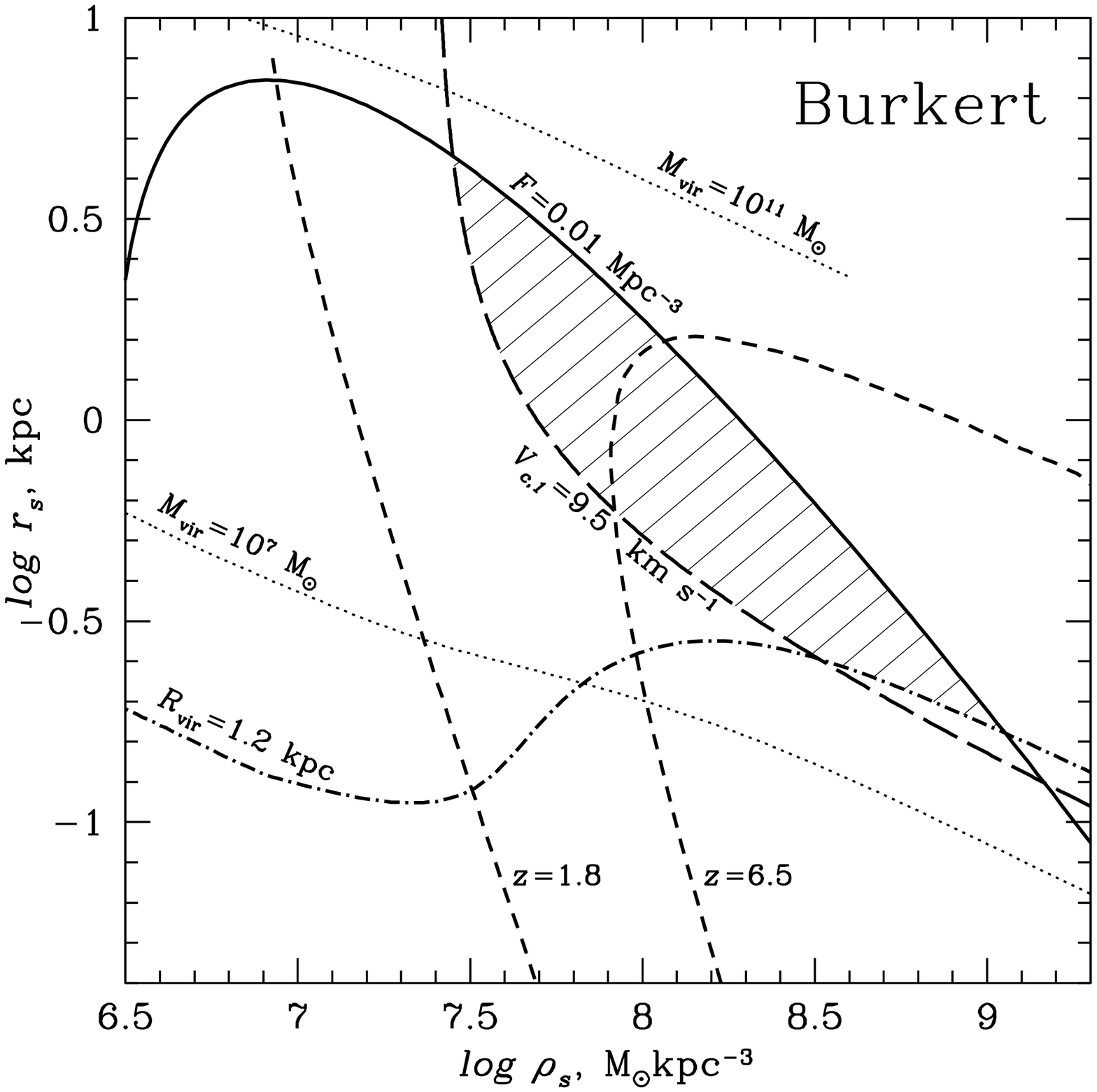}
\caption {Exclusion plots for Draco DM halo parameters $\rho_s$ and $r_s$. (Both NFW case,
left panel, and Burkert case, right panel, are shown.) The shaded areas
correspond to DM halos which satisfy all global constraints. Here $V_{c,1}$ is
the circular speed at the distance $r_1=0.74$~kpc from the center of the halo.
\label{fig1} }
\end{figure*}

It is interesting that one can derive quite general and (stellar)
model-independent constraints on the properties of the Draco DM halo by
combining the available observational data on this galaxy with the predictions
of cosmology. Throughout this paper we assume that the distance to Draco is
$82\pm 6$~kpc \citep{mat98}.  It is convenient to consider different DM halo
constraints in the plane of two scaling parameters, $\rho_s$ and $r_s$. We
summarize the global constraints in Figure~\ref{fig1}.

The most obvious constraint is that the Draco DM halo should have formed before
the bulk of the Draco stars formed. We assume that the first star burst in Draco
took place at least 10~Gyr ago \citep{mat98}, or at $z\geqslant 1.8$. In
Figure~\ref{fig1}, the areas to the right of the dashed lines marked ``$z=1.8$''
correspond to halos older than 10~Gyr.

The next constraint, $F\geqslant F_{\rm min}$, comes from the requirement for DM
halos to be abundant enough to explain the observed number ($\sim 20$) of dwarf
spheroidal galaxies in the Local Group. Following \citet{mcs05}, we adopt
$F_{\rm min}=0.01$~Mpc$^{-3}$.  As the function $F$ does not explicitly depend
on $\rho_s$ and $r_s$ (it depends on $m_{\rm vir}$ and $z$), in \citet{mcs05} we
designed a numerical method for finding the most probable combination of
($m_{\rm vir}$,$z$) corresponding to given values of ($\rho_s$,$r_s$). As a
by-product we also obtain the corresponding value of $\nu_c$. The areas below
the solid lines in Figure~\ref{fig1} correspond to DM halos which were abundant
enough to have been progenitors of dwarf spheroidals in the Local Group.

\begin{figure*}
\plottwo{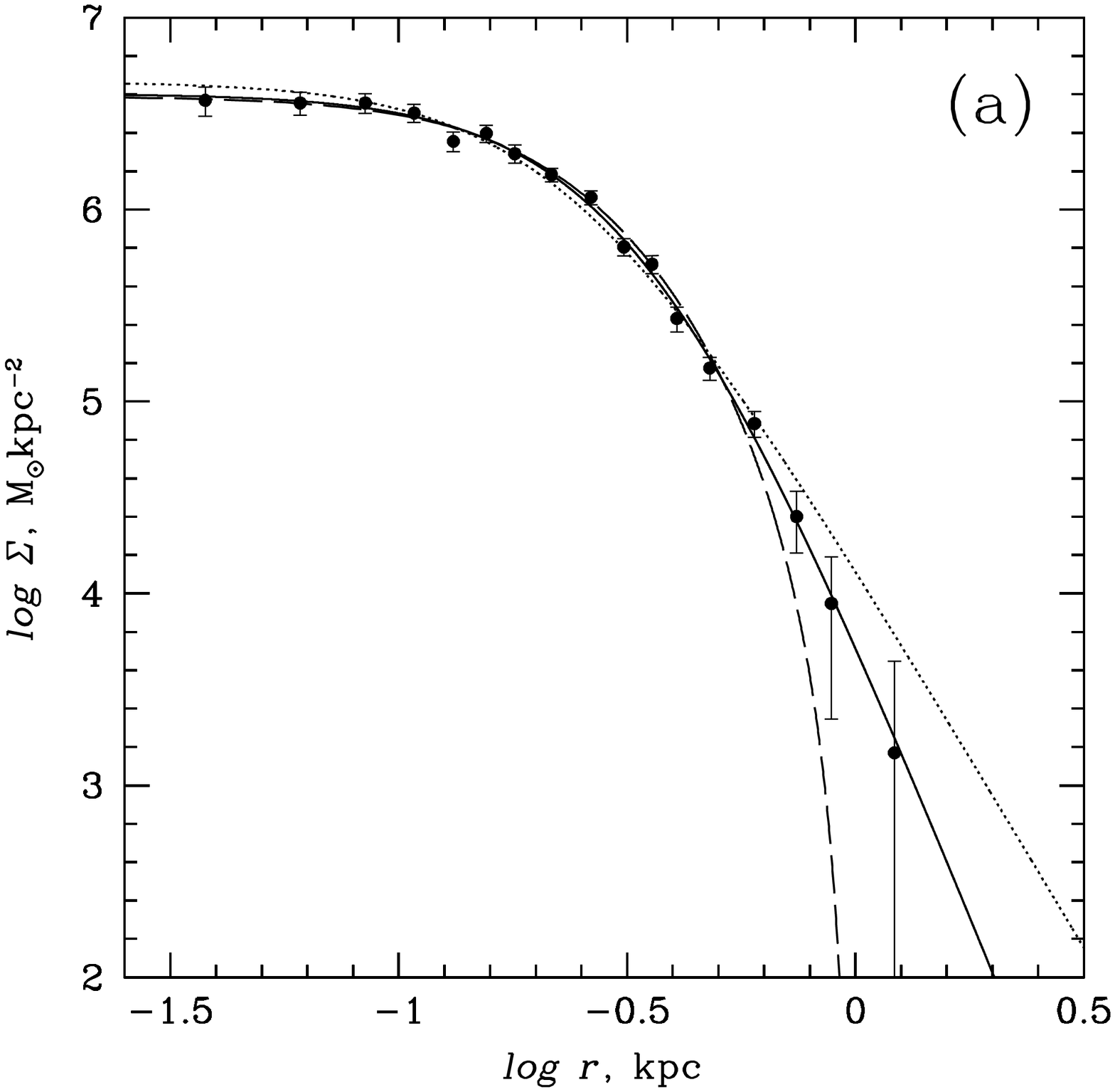}{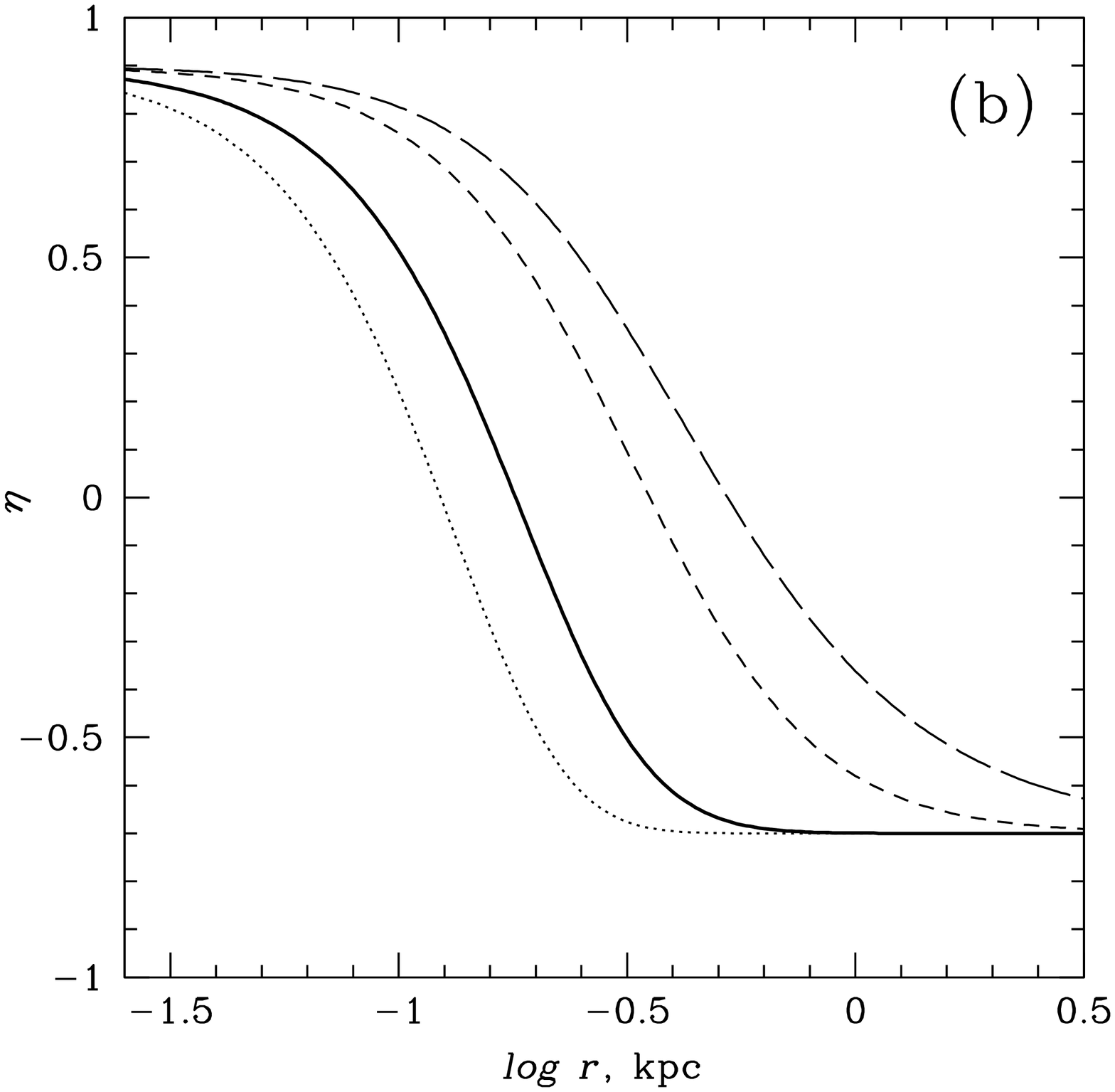}
\caption {(a) Stellar surface density profiles for Draco. The observed profile of 
\citet[their sample S2]{ode01} is shown as
solid circles with one-sigma error bars. The best fitting profiles for
Plummer-like model with $\alpha=7$, Plummer model, and King model are shown as
solid, dotted, and dashed lines, respectively.  (b) Stellar anisotropy profile
$\eta(r)$ for the best-fitting Draco model with NFW DM halo ($\alpha=7$, $\log
\rho_s=7.2$, $\log r_s=0.45$, $\lambda=1$, $\eta_0=0.9$, $\eta_1=-0.7$) is shown
as a solid line. For comparison, anisotropy profiles are shown for other values
of $\lambda$: 0.5 (dotted line), 3 (short-dashed line), and 5 (long-dashed line).
\label{fig2} }
\end{figure*}

Assuming that the tidal field of the Milky Way has not perturbed significantly
the stars in the outskirts of Draco out to a distance of $r_{\rm out}\sim
1.2$~kpc from its center (the last observed point in the Draco surface
brightness profile of \citealt{ode01}, see Figure~\ref{fig2}a), the third constraint can be written as
$r_{\rm vir}\geqslant r_{\rm out}$. (It is hard to imagine stars forming
beyond the virial radius of its DM halo). The halos in the areas above the
dash-dotted lines in Figure~\ref{fig1} satisfy the above criterion.

\begin{figure}
\plotone{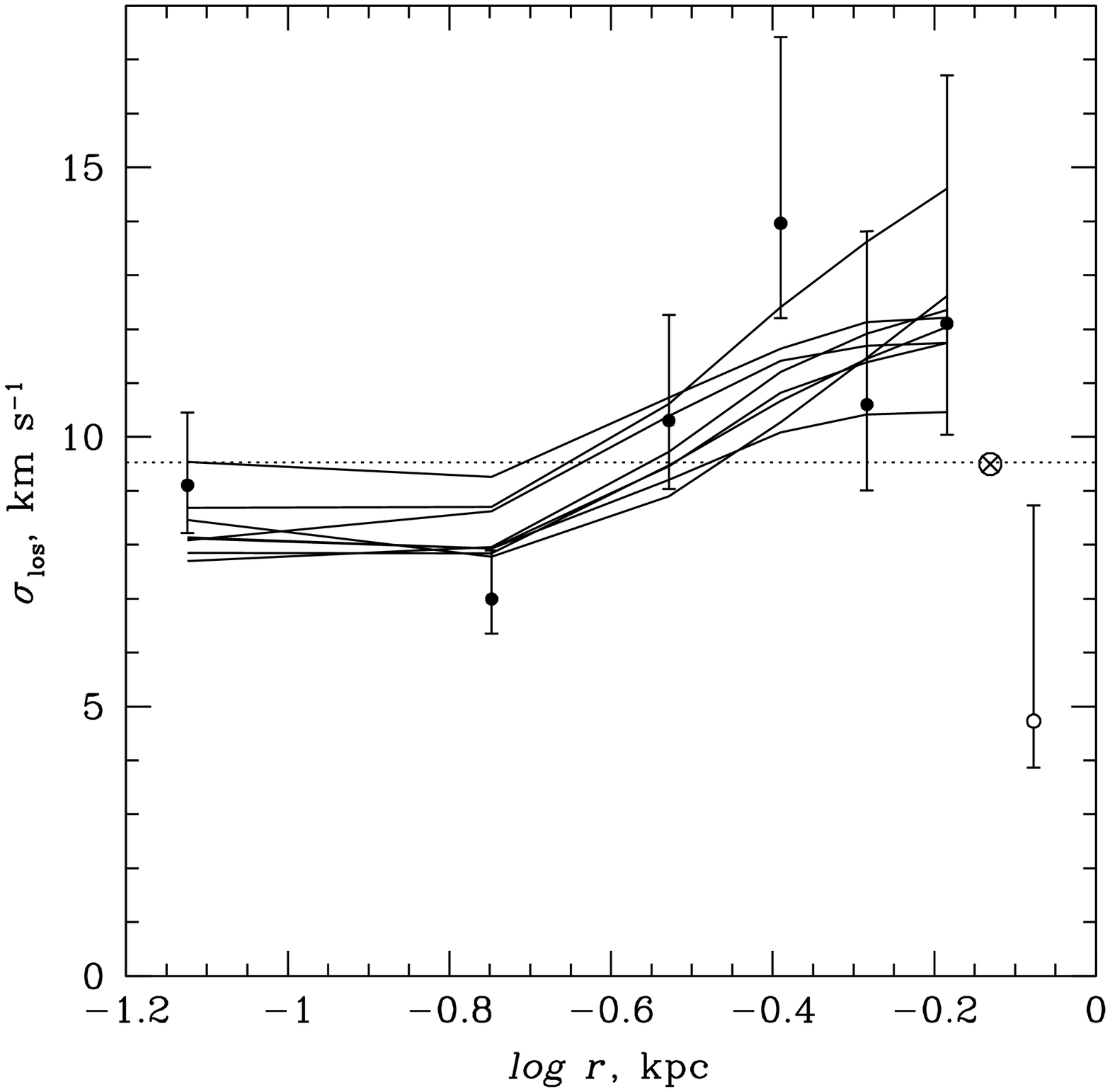}
\caption {Line-of-sight stellar velocity dispersion profile for Draco. Observational
data of \citet{wil04} are shown as circles with one-sigma error bars. The last
point which we consider to be unreliable  is shown as an empty circle with
error bars. The horizontal dotted line marks the global $\sigma_{\rm
los}=9.5$~km~s$^{-1}$ for Draco. The $\sigma_{\rm los}$ profiles for our 8
best-fitting stellar models are shown as solid lines. The big circle with a cross
shows our assumption for $\sigma_{\rm los}$ at the distance of $r_1=0.74$~kpc from the
Draco center.
\label{fig4} }
\end{figure}

In Figure~\ref{fig4} we plot the observed line-of-sight velocity dispersion
profile for Draco (from \citealt{wil04}). It has been noted that in Draco and
Ursa Minor the galactic outskirts appear to be kinematically cold
\citep{wil04}. There appears to be no good explanation for this phenomena. Given
the fact that in the case of Draco the only (outermost) radial bin which is ``cold''
contains only 6 stars with measured line-of-sight velocities, and is marginally
(at $\sim 1$ sigma level) consistent with the global average for Draco,
$\langle \sigma_{\rm los}\rangle=9.5$~km~s$^{-1}$, we decided to exclude the last bin from our
analysis. Our assumption is that at the distance of $r_1=0.74$~kpc from the Draco
center (half-way between the two last points in Figure~\ref{fig4}) the
line-of-sight velocity dispersion for Draco is roughly equal to the global
average\footnote{\parbox[t]{8cm}{This is consistent with the results of \citet{mun05} who observed Draco to have 
$\sigma_{\rm los}\sim 10$~km~s$^{-1}$ out to a distance of $\sim 1$~kpc from the center.}}:
 $\sigma_{\rm los}(r_1)\sim 9.5$~km~s$^{-1}$ (the circle with a cross in
Figure~\ref{fig4}). At this distance the stellar density in Draco is steeply
declining (see Figure~\ref{fig2}a). As a result, most of stars observed at the
projected distance $r_1$ from the Draco center are located roughly in the plane
of the sky at the {\it spatial} distance $r_1$ from the center of the dwarf.  We
can then use $\sigma_{\rm los}(r_1)$ as a lower limit of Draco circular velocity
at this distance, $V_{c,1}$. Indeed, if one considers the extreme of purely
radial orbits, the line-of-sight velocity dispersion at large distance from the
galactic center will be close to zero. In the opposite extreme of purely circular
orbits, one can show that $\sigma_{\rm los}$ becomes approximately equal to the
circular velocity at this distance. We plot the solution of the equation
$V_{c,1}=9.5$~km~s$^{-1}$ as long-dashed lines in Figure~\ref{fig1}. The areas
above these lines correspond to halos which satisfy the above criterion. The
result we derived here is approximate, but of model-independent nature. We
present more accurate (but also more model-dependent) treatment in
\S~\ref{st_model}, where we fit stellar models to all the reliable observed  $\sigma_{\rm los}$
points in Figure~\ref{fig4}.

The last constraint is that the virial mass of the Draco halo is somewhere
between ``ridiculously'' low and high values $10^7$ and $10^{11}$~M$_\odot$
(the area between two dotted lines in Figure~\ref{fig1}). The lower limit is
even lower, by a factor $2-3$, than the classical ``mass follows light''
estimates \citep{mat98,ode01}. The upper limit corresponds to a satellite which
would very quickly spiral in to the center of the Milky Way due to dynamical
friction. As one can see, the last constraint does not add any new information
to our exclusion plots in Figure~\ref{fig1}.

From comparison of Figures~\ref{fig1}a and \ref{fig1}b one can see that all the
constraints except for the fourth one ($V_{c,1}\geqslant 9.5$~km~s$^{-1}$) are
very similar for both NFW and Burkert DM density profiles. This is intimately
linked to our assumption that for a given virial mass $m_{\rm vir}$ and
redshift $z$ both types of halos have the same concentration -- given by
equation~(\ref{eqc}). This automatically makes the scaling radii $r_s$ equal in
both cases. At the same time, one can show that the scaling densities for NFW
and Burkert halos are related through

\begin{equation}
\frac{\rho_{s,\rm Burkert}}{\rho_{s,\rm NFW}} = 2\frac{\ln(1+c)-c/(1+c)}{\ln(1+c)+[\ln(1+c^2)]/2-\arctan c},
\end{equation}

\noindent which is close to unity for realistic halos (the ratio changes
from $0.966\dots 0.921$ for $c=3.5\dots 10$). The fourth constraint, unlike
the rest of the criteria, does not deal with a global property of a halo (such as
$m_{\rm vir}$ and $z$ or their derivatives -- $r_{\rm vir}$ and $F$). Instead,
it deals with the average DM density within a certain fixed radius -- which can
be dramatically different for cuspy NFW and flat-cored Burkert models.

The shaded areas in Figure~\ref{fig1} correspond to DM halos which satisfy all
of the above global constraints. As one can see, the Draco halo parameters are
not particularly well constrained, especially in the case of the NFW profile. Still, one
can make a few interesting observations. Firstly, the most restrictive (and useful)
constraints are $F\geqslant 0.01$~Mpc$^{-3}$ and $V_{c,1}\geqslant
9.5$~km~s$^{-1}$. Secondly, there is plenty of room for Draco to have formed
before the end of the reionization of the universe at $z\sim 6.5$ (shaded areas to
the right of the dashed lines marked ``$z=6.5$'' in Figure~\ref{fig1}). Thirdly,
we can derive the range of possible values of different halo parameters. For NFW
halos, our exclusion plot implies that $\log m_{\rm vir}=7.7\dots 10.7$,
$z=1.8\dots 11$, $r_s=0.16\dots 7.8$~kpc, and $\log \rho_s =7.0\dots 9.1$. (Here
units for $m_{\rm vir}$ and $\rho_s$ are M$_\odot$ and M$_\odot$~kpc$^{-3}$,
respectively.) For Burkert halos, $\log m_{\rm vir}=7.8\dots 10.5$, $z=3.9\dots
11$, $r_s=0.17\dots 5.0$~kpc, and $\log \rho_s =7.4\dots 9.1$. Interestingly,
for both NFW and Burkert cases, Draco could not have formed before $z\sim 11$,
or more than 13.2~Gyrs ago. The reason for that is that at larger redshifts
DM halos with the virial radii $\geqslant 1.2$~kpc become too rare to correspond to a typical
dwarf spheroidal galaxy in the Local Group.

The main purpose of generating the exclusion plots in Figure~\ref{fig1} was to
significantly reduce the computational burden in the next step of our analysis,
described in the next section, where we use the whole observed line-of-sight velocity
dispersion profile along with the surface brightness profile to find the 
best-fitting stellar models and to further reduce the uncertainty in ($\rho_s$,$r_s$)
values for the Draco DM halo.

\section{Stellar Model}
\label{st_model}

The equilibrium state of a spherically symmetric stellar system can
described by the Jeans equation,

\begin{equation}
\label{Jeans}
\frac{1}{\rho_*}\frac{d(\rho_* \sigma_r^2)}{dr} + \frac{2}{r}\left(\sigma_r^2 - \sigma_t^2\right)
= -\frac{d\Phi}{dr}
\end{equation}

\noindent \citep{bin87}, which is obtained by taking the first velocity moment of the collisionless Boltzmann equation.
Here $r$ is the distance from the center of the system, $\rho_*$ is the stellar
density, $\sigma_r$ and $\sigma_t$ are the one-dimensional stellar radial and
tangential velocity dispersions, respectively, and $\Phi$ is the total
gravitational potential (due to stars and DM). The radial gradient of the
gravitational potential can be calculated as $d\Phi/dr=G[m(r)+m_*(r)]/r^2$,
where $m(r)$ and $m_*(r)$ are the enclosed DM and stellar masses, respectively,
and $G$ is the gravitational constant.  From equations (\ref{eq1}--\ref{eq2}) we
derived

\begin{eqnarray}
m(r)&=&4\pi r_s^3 \rho_s [\ln(1+x)-x/(1+x)]  \hspace{1.2cm} \mbox{(NFW)},\label{m_DM}\\
m(r)&=&2\pi r_s^3 \rho_s [\ln(1+x)+(1/2)\ln(1+x^2)-\arctan x]\nonumber \\  
&&\hspace{5cm}\mbox{(Burkert).}\label{m_DM2}
\end{eqnarray}

\noindent Here $x\equiv r/r_s$. We neglect impact of baryons on DM distribution,
as in Draco the stellar density is more than an order of magnitude lower than
DM density even at the center of the galaxy.

The Plummer density profile,

\begin{equation}
\label{rho_*}
\rho_*=\rho_0 \left[ 1 + (r/b)^2\right]^{-5/2}
\end{equation}

\noindent \citep{bin87}, is used sometimes to describe simple spherically symmetric stellar systems,
such as globular clusters and dwarf spheroidal galaxies. It has a core of
size $b$ and a power-law envelope with the slope $\gamma=-5$. We found that a
``generalized Plummer law'',

\begin{equation}
\label{genplum}
\rho_*=\rho_0 \left[ 1 + (r/b)^2\right]^{-\alpha/2},
\end{equation}

\noindent which has a surface density profile of the form

\begin{equation}
\label{Sigma}
\Sigma=\Sigma_0 \left[ 1 + (R/b)^2\right]^{-(\alpha-1)/2},
\end{equation}

\noindent provides much better fit to the Draco star count
profile of \citet{ode01} than the Plummer model and the theoretical
\citet{K66} model, if one chooses $\alpha=7$ (see Figure~\ref{fig2}a)
\footnote{For the Draco star count of \citet{wil04} the best-fitting value
of $\alpha$ is 6.}.  Here $\alpha$ is an integer number $\geqslant 2$ and $R$ is
the projected distance from the center of the system. For $\alpha
\geqslant 4$, the total stellar mass can be calculated as

\begin{equation}
M_*=\frac{2\pi \Sigma_0 b^2}{\alpha-3}.
\end{equation}

\noindent For $\alpha=7$, the stellar enclosed mass is

\begin{equation}
\label{m_*}
m_*(r)=\frac{4\pi \rho_0 r^3}{15} \frac{5+2(r/b)^2}{(1+r^2/b^2)^{5/2}},
\end{equation}

\noindent and $\rho_0=15\Sigma_0/(16 b)$. The $\chi^2$ fitting of
equation~(\ref{Sigma}) to the Draco profile of
\citet{ode01} gave $\rho_0=1.08\times 10^7$~M$_\odot$~kpc$^{-3}$ and
$b=0.349$~kpc. (We assumed that the stellar $V$-band mass-to-light ratio of
Draco stars is $\Upsilon=1.32$, which is an average value of the Salpeter and
composite model estimates of \citealt{mov98}.)

Traditionally, in equation~(\ref{Jeans}) one uses $\sigma_r$ and $\beta\equiv
1-\sigma_t^2/\sigma_r^2$ instead of $\sigma_r$ and $\sigma_t$. The anisotropy
parameter $\beta$ is equal to $-\infty$, 0, and 1 for purely tangential
(circular), isotropic, and purely radial stellar orbits. We advocate a different
anisotropy parameter,

\begin{equation}
\eta\equiv \frac{\sigma_r^2-\sigma_t^2}{\sigma_r^2+\sigma_t^2}.
\end{equation}

\noindent Unlike $\beta$, parameter $\eta$ is symmetric: it is equal to $-1$,
0, and 1 for circular, isotropic, and purely radial orbits. The equations
connecting $\beta$ and $\eta$ are $\eta=\beta/(2-\beta)$ and
$\beta=2\eta/(1+\eta)$. Another useful expression is $\sigma_t^2=\sigma_r^2
(1-\eta)/(1+\eta)$. The Jeans equation~(\ref{Jeans}) can now be rewritten as

\begin{equation}
\label{Jeans2}
\frac{1}{\rho_*}\frac{d(\rho_* \sigma_r^2)}{dr} + \frac{4\eta}{1+\eta}\frac{\sigma_r^2}{r}
= -\frac{d\Phi}{dr}.
\end{equation}

As there are two unknown functions in equations~(\ref{Jeans}) or (\ref{Jeans2}),
$\sigma_r(r)$ and $\sigma_t(r)$ (or $\beta[r]$, or $\eta[r]$), one customarily
assumes the shape of the anisotropy profile and then solves the ordinary
differential equation for $\sigma_r(r)$. The boundary condition is
$\sigma_r(r\rightarrow\infty)=0$. Traditional choices for the anisotropy profile
are (1) $\beta=constant$ (with the special case of $\beta=0$, or isotropic
stellar orbits), and (2) the Osipkov-Merritt profile \citep{osi79,mer85},

\begin{equation}
\beta=\left[ 1+\left(r_a/r\right)^2\right]^{-1}.
\end{equation}

\noindent In the latter case, the stellar system is isotropic at the center,
reaches $\beta=0.5$ at $r=r_a$, and becomes purely radially anisotropic in
infinity. Unfortunately, the two above choices are very limited. We propose
instead a much more flexible anisotropy profile,

\begin{equation}
\label{eta}
\eta = \eta_0 + (\eta_1-\eta_0)\left[1-\left(\rho_*/\rho_0\right)^{1/\lambda}\right],
\end{equation}

\noindent which is applicable to systems with a flat core (such as generalized Plummer model
or King model). Here $\lambda$ is a positive number of order of unity, and $\eta_0$
and $\eta_1$ are asymptotic values of the anisotropy parameter $\eta$ for $r\rightarrow 0$
and $r\rightarrow\infty$, respectively. As one can see, the profile in
equation~(\ref{eta}) does not explicitly depend on $r$, like the
Osipkov-Merritt profile. Instead, it depends on the stellar density $\rho_*$. We believe
it is a reasonable approach, as in many realistic stellar systems the same
dynamical processes shape simultaneously both density and anisotropy profiles.
 The examples are the collapse of initially homogeneous warm
stellar sphere \citep{alb82,mas05a} and the expansion of a newly formed stellar
system in a spherical galaxy with a DM halo after removal of the leftover gas by the feedback mechanisms
\citep{mcs05}. In both cases, the stellar orbits become increasingly radially
anisotropic in the outskirts of the relaxed system, where the density is steeply
declining.

Both $\beta=constant$ and the Osipkov-Merritt profiles can be considered as
special cases of our more general expression in equation~(\ref{eta}). Indeed,
fixing $\eta_0=\eta_1$ would correspond to the case of $\beta=constant$; using
the generalized Plummer density profile from equation~(\ref{genplum}) and
setting $\lambda=\alpha/2$ produces the Osipkov-Merritt profile with
$r_a=b/\sqrt{2}$.

The parameter $\lambda$ in equation~(\ref{eta}) controls how sensitive the
anisotropy parameter is to changes in density. To illustrate this effect, in
Figure~\ref{fig2}b we show anisotropy profiles for Draco with $\eta_0=0.9$,
$\eta_1=-0.7$, and $\lambda=0.5,1,3,5$. One can see that by varying $\lambda$ by a
factor of a few a large range of possible anisotropy profiles is produced.

We designed a numerical algorithm to find values of the parameters controlling
the shape of the anisotropy function $\eta$ ($\eta_0$, $\eta_1$, and $\lambda$)
which would $\chi^2$ minimize the deviation of the simulated line-of-sight
velocity dispersion profile from the observed one (circles with error bars in
Figure~\ref{fig4} --- we excluded from our analysis the last point as an
unreliable one) for given $\rho_s$, $r_s$, and the halo profile (NFW or Burkert).
The procedure consists of the following six steps.

\begin{figure*}
\plottwo{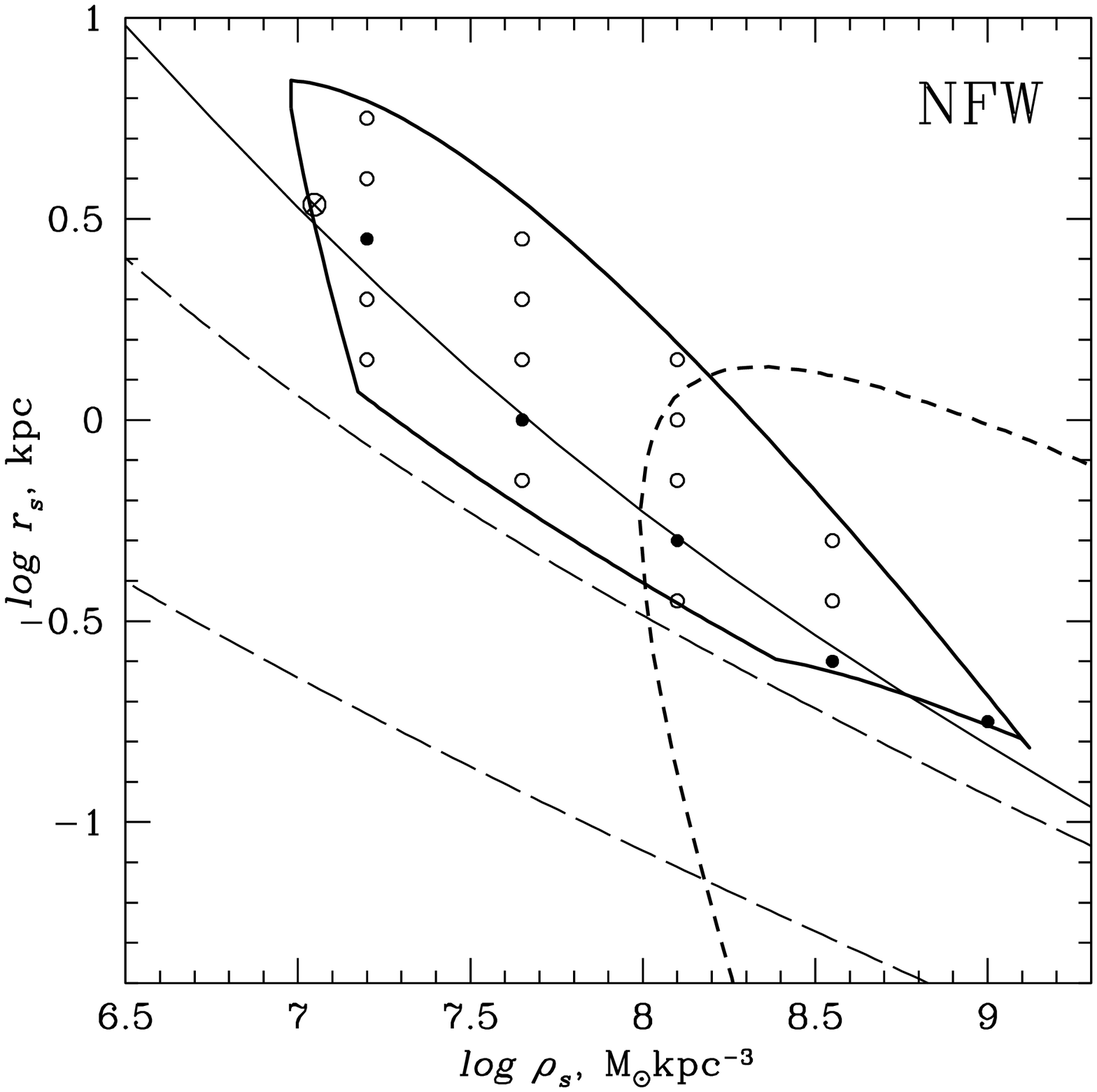}{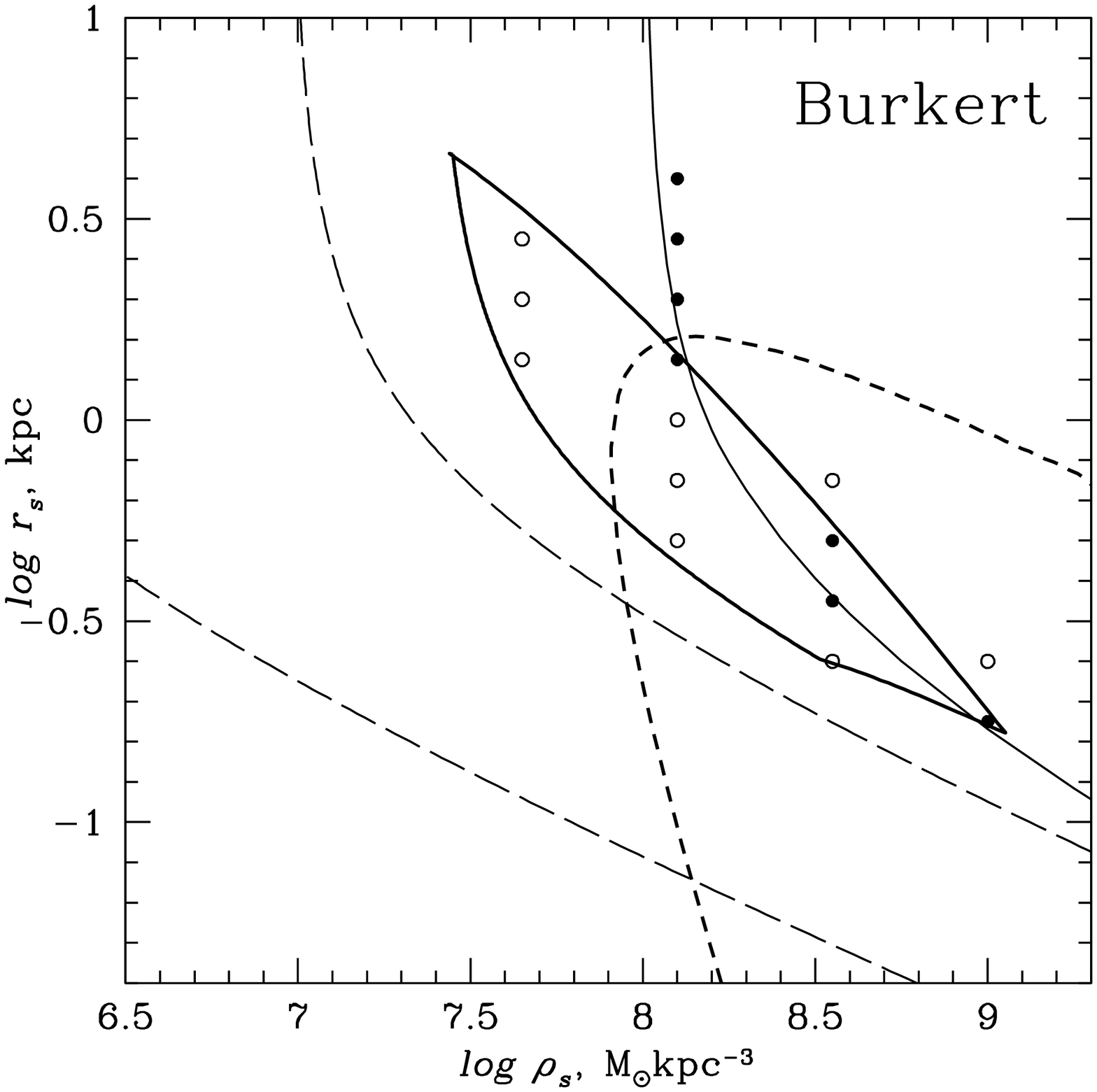}
\caption {Best-fitting stellar models for different Draco DM halo parameters $\rho_s$ and 
$r_s$. Solid (empty) circles correspond to models which have $\chi^2<9.5$
($\chi^2>9.5$) between the modeled and observed line-of-sigh velocity
dispersion profiles. Thick solid lines mark the areas where all the global
constraints on ($\rho_s$,$r_s$) are satisfied (see Figure~\ref{fig1}). Thin
solid lines correspond to isothermal stellar models with
$\sigma=9.5$~km~s$^{-1}$. Short-dashed lines correspond to halos formed at
$z=6.5$.  Long-dashed lines correspond to halos with the analytical tidal radius equal to 0.85~kpc for the
extreme values of the orbital pericentric distance, $R_p=2.5$ and $70$~kpc (see
\S~\ref{Orbits}).  Circle with a cross marks the best-fitting Draco model from \citet{mcs05} for
the Draco surface brightness profile of \citet{ode01}.
\label{fig3} }
\end{figure*}

(1) We choose values of $\log \rho_s$ and $\log r_s$ from a grid with spacings
of 0.45~dex and 0.15~dex, respectively. (The values of the increments were
chosen to lead to a factor of two increase in the halo virial mass.) The
reference point of the grid is $\log \rho_s =9$ and $\log r_s =0$. Only those
grid points are considered which are lying within the shaded zones in
Figure~\ref{fig1}. In the Burkert case, we also considered a few points lying
outside of the shaded area in attempt to bracket the point with the best
$\chi^2$. All these grid points are shown as circles (either empty or filled)
in Figure~\ref{fig3}. Overall, we considered 35 different DM halo models.

(2) For each of the DM halo models, we consider 1764 different combinations of
anisotropy shape parameters $\eta_0=-1,-0.9,-0.8,\dots,1$,
$\eta_1=-1,-0.9,-0.8,\dots,1$, and $\lambda=0.5,1,3,5$. For each combination, we
solve the Jeans equation~(\ref{Jeans2}) numerically with the anisotropy, stellar
density, enclosed DM mass, and enclosed stellar mass profiles given by
equations~(\ref{eta}), (\ref{rho_*}), (\ref{m_DM}--\ref{m_DM2}), and (\ref{m_*}),
respectively. We adopt $\alpha=7$, $\rho_0=1.08\times
10^7$~M$_\odot$~kpc$^{-3}$, and $b=0.349$~kpc. The solution of the Jeans equation is the radial
velocity dispersion profile $\sigma_r(r)$.

(3) For each of the above $\sim 60,000$ models we generate a spherically
symmetric $N$-body stellar model, with the number of particles
$N=10,000$. Stellar particles are distributed randomly, with the density profile
given by equation~(\ref{rho_*}). Each stellar particle is assigned random values
of the radial and two tangential components of the velocity vector: $V_r$,
$V_\theta$, and $V_\phi$. All three components are assumed to have a Gaussian
distribution, with the dispersions $\sigma_r(r)$, $\sigma_t(r)$, and
$\sigma_t(r)$, respectively. This is an approximate method of generating a
close-to-equilibrium $N$-body system with an arbitrary density and anisotropy
profiles.  The accurate method would involve numerically calculating the
distribution function, which is a very computationally expensive procedure. This
would render our approach unfeasible.

(4) We calculate the line-of-sight velocity dispersion profile for the generated
$N$-body stellar models integrated over the same projected radial bins as in the
observed profile of \citet[]{wil04} (the edges of their bins are 0,0.12,0.24,0.36,0.48,0.60,0.72~kpc --
Jan~Kleyna, private communication). For this we use the projection method of
\citet[Appendix B]{mas05a} where we explicitly use the spherical symmetry of our
stellar system.

(5) We calculate $\chi^2$ difference between the observed line-of-sight velocity
dispersion profile (the six reliable points in Figure~\ref{fig4}) and the modeled
one. As the observational error bars are asymmetric, for calculating $\chi^2$ we
use the appropriate one-sided value of the standard deviation depending on which side of the
observed point the model point is located.

(6) For each DM halo model, we choose one of 1764 models, with different
$\eta_0$, $\eta_1$, and $\lambda$, which produces the lowest value of $\chi^2$.

For most DM halo models, we could not find a good fit to the observed
line-of-sight velocity dispersion profile (with some models having
$\chi^2>100$): all modeled $\sigma_{\rm los}(r)$ points were either well above
or well below the observed ones. Only a few models produced $\chi^2<9.5$ (solid
circles in Figure~\ref{fig3}). It is interesting that for both NFW and Burkert
cases the best-fitting models follow very closely a sequence of isothermal
stellar models with the total one-dimensional stellar velocity dispersion
$\sigma_{\rm tot}=[(\sigma_r^2+2\sigma_t^2)/3]^{1/2}=9.5$~km~s$^{-1}$ (thin
solid lines in Figure~\ref{fig3}). We obtained the isothermal solutions by solving
the appropriate Jeans equation,

\begin{equation}
\label{Jeans3}
\frac{1}{\rho_*}\frac{d(\rho_* \sigma_r^2)}{dr} + \frac{3}{r}\left(\sigma_r^2 - \sigma_{\rm tot}^2\right)
= -\frac{d\Phi}{dr},
\end{equation}

\noindent with the boundary condition $\sigma_r(0)=\sigma_{\rm tot}$. For isothermal
systems the usual Jeans equation requirement $\sigma_r(r\rightarrow\infty)=0$
does not hold in general case, and at some radius $r_{\rm max}$ the solution
breaks down when $\sigma_r^2$ becomes negative. The isothermal model lines shown
in Figure~\ref{fig3} were obtained by finding the value of $r_s$ which would
maximize $r_{\rm max}$ for given $\rho_s$. Typically $r_{\rm max}\gg 1$~kpc, and
only for the NFW model with $\rho_s=10^9$~M$_\odot$~kpc$^{-3}$ did
it drop down to 0.6~kpc.

The closeness of our best-fitting models to isothermal models does not imply
that the isothermal models are the best ones. We calculated $\chi^2$ differences
between the observed and modeled line-of-sight velocity dispersion profiles for
a few isothermal models, and they were substantially worse than for our
best-fitting models. We also plotted $\sigma_{\rm tot}(r)$ profiles for the
best-fitting models, and they were not isothermal. The explanation for the
closeness of our best-fitting models to the isothermal ones is in the virial
theorem. Given that the line-of-sight velocity dispersion profile and the
surface brightness profile are fixed, the virial theorem implies that all
realistic models should stay close to a certain line in the $(\rho_s,r_s)$
plane.

From Figure~\ref{fig3} one can see that the best-fitting models are well
bracketed inside the zones were all the global constrains on $\rho_s$ and $r_s$
from \S~\ref{global} are satisfied. In other words, we have a good agreement
between the global constraints and the detailed $\sigma_{\rm los}(r)$ analysis
constraints, which were derived in a very different fashion.  The exception is
the Burkert halos with $\log\rho_s=8.1$, where $\chi^2$ is slightly improving
when moving upwards into the area with $F< 0.01$~Mpc$^{-3}$. This is explained
by the fact that at this value of $\log\rho_s$ the isothermal sequence is making
a sharp upward turn.

\begin{table*}
\begin{center}
\caption{Best-fitting stellar models\label{tab1}} 
\begin{tabular}{cccccc|cccccccccc}
\tableline
\multicolumn{6}{c|}{Input Parameters} & \multicolumn{10}{c}{Derived Parameters}\\
\tableline
Model&Halo & $\log\rho_s$          & $\log r_s$ & $N_{\rm DM}$ & $\varepsilon_{\rm DM}$  & $\lambda$ & $\eta_0$ & $\eta_1$ & $\chi^2$ & $m_{\rm vir}$ & $z$ & $c$ & $r_{\rm vir}$ & $\nu_c$ & $\log F$   \\
     && M$_\odot$~kpc$^{-3}$ & kpc        & & pc &       &          &          &          &  M$_\odot$    &     &     &   kpc         &         &  Mpc$^{-3}$\\
\tableline
N1&    NFW      &     7.20  &     0.45  &$10^6$           &60&        1  &      0.9  &    $-0.7$ &      5.4   & $4.6\times 10^9$ & 2.46 & 5.55 & 15.6 & $-0.68$ & $-1.09$ \\
N2&    NFW      &     7.65  &     0.00  &$3.16\times 10^5$&28&        1  &      0.7  &    $-0.8$ &      6.6   & $5.2\times 10^8$ & 4.50 & 4.80 & 4.80 & $-0.23$ & $-0.15$ \\
N3&    NFW      &     8.10  &   $-0.30$ &$10^5$           &36&        1  &      0.8  &    $-0.9$ &      6.1   & $1.8\times 10^8$ & 7.10 & 4.55 & 2.28 &  0.54   & 0.08    \\
N4&    NFW      &     8.55  &   $-0.60$ &$10^5$           &20&        1  &      0.6  &    $-1.0$ &      8.9   & $6.9\times 10^7$ & 9.45 & 5.14 & 1.29 &  1.47   & $-0.11$  \\
N5&    NFW      &     9.00  &   $-0.75$ &$10^5$           &19&        1  &      1.0  &    $-1.0$ &      8.6   & $8.4\times 10^7$ & 10.7 & 6.92 & 1.23 &  2.80   & $-1.62$ \\
B1&    Burkert  &     8.10  &     0.15  &$10^6$           &37&        1  &      0.7  &      0.0  &      6.6   & $4.5\times 10^9$ & 6.74 & 4.98 & 7.03 &  1.48   & $-1.93$ \\
B2&    Burkert  &     8.55  &   $-0.45$ &$3.16\times 10^5$&30&        1  &      0.7  &    $-0.6$ &      6.5   & $2.1\times 10^8$ & 9.64 & 5.17 & 1.84 &  1.82   & $-0.97$ \\
B3&    Burkert  &     9.00  &   $-0.75$ &$10^5$           &19&        1  &      0.8  &    $-0.9$ &      6.2   & $9.1\times 10^7$ & 11.0 & 6.91 & 1.23 &  2.90   & $-1.83$ \\
\tableline
\end{tabular}
\tablecomments{Here $N_{\rm DM}$ and $\varepsilon_{\rm DM}$ are the number of the DM particles and the gravitational
softening length for DM particles in $N$-body simulations (see \S~\ref{isol}).}
\end{center}
\end{table*}

We list the parameters for the best-fitting models in Table~\ref{tab1}. We show
only the models with $\chi^2<9.5$ located within the globally constrained zones.
As for the Burkert halos with $\log\rho_s=8.55$ we have two comparably good
models, we chose the one which is closer to the isothermal sequence.

Analysis of Table~\ref{tab1} shows that a comparably good $\sigma_{\rm los}(r)$
fit can be obtained for a very large range in virial masses, from $\sim 7\times
10^7$ to $\sim 5\times 10^9$~M$_\odot$, for both NFW and Burkert halos. Despite
very large difference in DM masses and density profiles, all the best-fitting
stellar models have comparable values of the anisotropy parameters: $\lambda =1$
(in other words, anisotropy $\eta$ is a linear function of the stellar density),
$\eta_0\sim 0.8$, and $\eta_1\sim 0\dots -1$.

In Figure~\ref{fig4} we show the line-of-sight velocity dispersion profiles for
the 8 best-fitting models from Table~\ref{tab1} integrated over the same
projected radial bins as the observed profile. All the models correctly
reproduce the observational trend of a slight increase in $\sigma_{\rm los}$ 
with radius. However, two models with the largest virial mass (both NFW and Burkert) produce
profiles which are rising rather steeply at the last measured point. This
could present a problem if the result of \citet{mun05}, that the $\sigma_{\rm los}$ 
profile in the outskirts of Draco is almost flat, is confirmed with a larger
sample of stars. For less massive models, the line-of-sight velocity dispersion
profiles are levelling off at the last measured point, which is more in line with the results
of \citet{mun05}.

Assuming that the best-fitting stellar models follow closely the isothermal
sequence, from Figure~\ref{fig3} we can derive new (slightly better than in the
previous section) constraints on Draco's DM halo parameters: $\log m_{\rm
vir}\simeq 7.9\dots 9.7$ (for both NFW and Burkert halos), $z=1.8\dots 10$,
$r_s=0.21\dots 3.1$~kpc, $\log \rho_s =7.0\dots 8.8$ (for NFW halos), and
$z=6.8\dots 11$, $r_s=0.18\dots 1.4$~kpc, $\log \rho_s =8.1\dots 9.0$ (for
Burkert halos). The interesting result is that if {\it cosmological halos have a
Burkert-like flat-cored DM density profiles, then Draco should have formed
before the end of the reionization of the universe at $z\sim 6.5$}.

\section{Evolution in the Milky Way Potential}

\subsection{Possible Orbits in the Galactic Potential}
\label{Orbits}

To carry out tidal stripping simulations for Galactic satellites, it is of
principal importance to know reasonably well the shape of the gravitational
potential of the Milky Way. One popular Milky Way model often used to calculate
orbits of Galactic globular clusters and dwarf spheroidals is that of
\citet{joh99}. This model consists of three components: (1) a disk represented by
\citet{miy75} potential with the mass $1.0\times 10^{11}$~M$_\odot$,
radial scale-length 6.5~kpc, and scale-height 0.26~kpc, (2) a spherical bulge
with a \citet{her90} potential, mass $3.4\times 10^{10}$~M$_\odot$, and
scale-length 0.7~kpc, and (3) an isothermal halo with $\sigma=128$~km~s$^{-1}$
and a flat core of size 12.0~kpc. The model was designed to reproduce the
observed flat Galactic rotation curve between 1 and 30~kpc.

There are two major disadvantages of the above model for our work. Firstly, it
is axisymmetric, which adds an additional degree of freedom to our already
multi-dimensional problem. (Orbits in axisymmetric potentials depend on all
three components of the space velocity vector of the satellite, whereas in
spherical potentials orbits depend only on the radial, $V_r$, and tangential,
$V_t$, space velocity components.) Secondly, the DM halo of \citet{joh99} is an
isothermal sphere with the rotation curve asymptotically approaching
$V_c=181$~km~s$^{-1}$ as $r\rightarrow\infty$, whereas in the cosmological halos
the rotation curves are declining in the outskirts. 

We decided to use a simple static NFW potential,

\begin{equation}
\label{pot}
\Phi=-4\pi G\varrho_s R_s^3 \ln(1+R/R_s) / R,
\end{equation}

\noindent as the Milky Way model for our project. The scaling radius $R_s$ and density $\varrho_s$
of an NFW halo can be determined from the virial mass $M_{\rm vir}$ and
concentration $C$.  These quantities are still not very well known for the Milky
Way. Traditionally, one uses different Galactic objects (stars, gas, globular
clusters, dwarfs spheroidals) with known line-of-sight velocity and, in some
cases, proper motion as kinematical tracers of the Galactic potential, with
different authors obtaining quite different results. The favored model of
\citet*{kly02}, for example, has a virial mass of $10^{12}$~M$_\odot$ and
concentration $10\dots 17$. Other recent NFW halo based models have $M_{\rm
vir}=(0.7-1.7)\times 10^{12}$~M$_\odot$, $C=5\dots 12$ \citep{car05}, and
$M_{\rm vir}=(0.6-2.0)\times 10^{12}$~M$_\odot$, $C=18$ \citep{bat05}.  Non-NFW
models of \citet{sak03} give larger values for the total mass of the Milky Way:
$M_{\rm vir}=(1.5-3.0)\times 10^{12}$~M$_\odot$ if Leo~I is gravitationally
bound to the Galaxy, and $M_{\rm vir}=(1.1-2.2)\times 10^{12}$~M$_\odot$ if not.

We adopted an intermediate value for the Galactic virial mass, $M_{\rm
vir}=1.5\times 10^{12}$~M$_\odot$.  The median concentration of cosmological
halos with such mass at $z=0$ is $C=13.2$ \citep{bul01}. The halo scaling
parameters are then $R_s=22.6$~kpc and $\varrho_s=6.0\times
10^6$~M$_\odot$~kpc$^{-3}$, and the virial radius is $R_{\rm vir}=298$~kpc.

\begin{figure}
\plotone{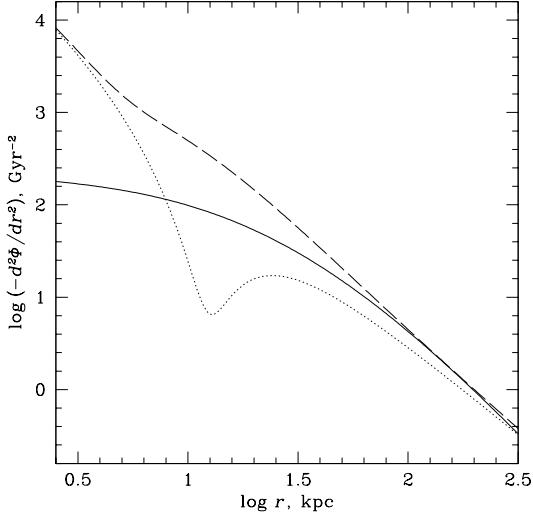}
\caption {Radial tidal acceleration profiles for the Milky Way models.
Solid line correspond to our NFW model. The composite model of \citet{joh99} is
shown as dashed (in the Galactic plane) and dotted (in the polar direction)
lines.
\label{fig7} }
\end{figure}

The radial tidal acceleration $-d^2\Phi/dr^2$ of
our NFW halo is comparable to that of the composite model of \citet{joh99}.  As
one can see in Figure~\ref{fig7}, the tidal acceleration profile for the NFW
halo is located between the two extreme profiles for the
\citet{joh99} model (in the Galactic plane and along the polar axis) down to a
radius of $\sim 10$~kpc.

We obtained the pericentric and apocentric distances, $R_p$ and $R_a$, for
closed Draco's orbits in the potential given by equation~(\ref{pot}) by solving
numerically the following non-linear equation
\citep{bin87}:

\begin{equation}
\frac{1}{R^2}+\frac{2[\Phi(R)-\Phi(R_0)]-V_r^2-V_t^2}{R_0^2 V_t^2} = 0,
\end{equation}

\noindent where $R_0$, $V_r$, and $V_t$ are the Draco's current distance from the Galactic center,
radial velocity, and tangential velocity, respectively. If proper motion is
known, the two space velocity vector components $V_r$ and $V_t$ can be
calculated using the procedure outlined in Appendix.  The radial
orbital period is obtained by solving numerically the following integral
\citep{bin87}:

\begin{equation}
P=2\int_{R_p}^{R_a}\frac{dR}{\sqrt{V_r^2+V_t^2-2[\Phi(R)-\Phi(R_0)]-R_0^2V_t^2/R^2}}.
\end{equation}

\begin{figure*}
\plottwo{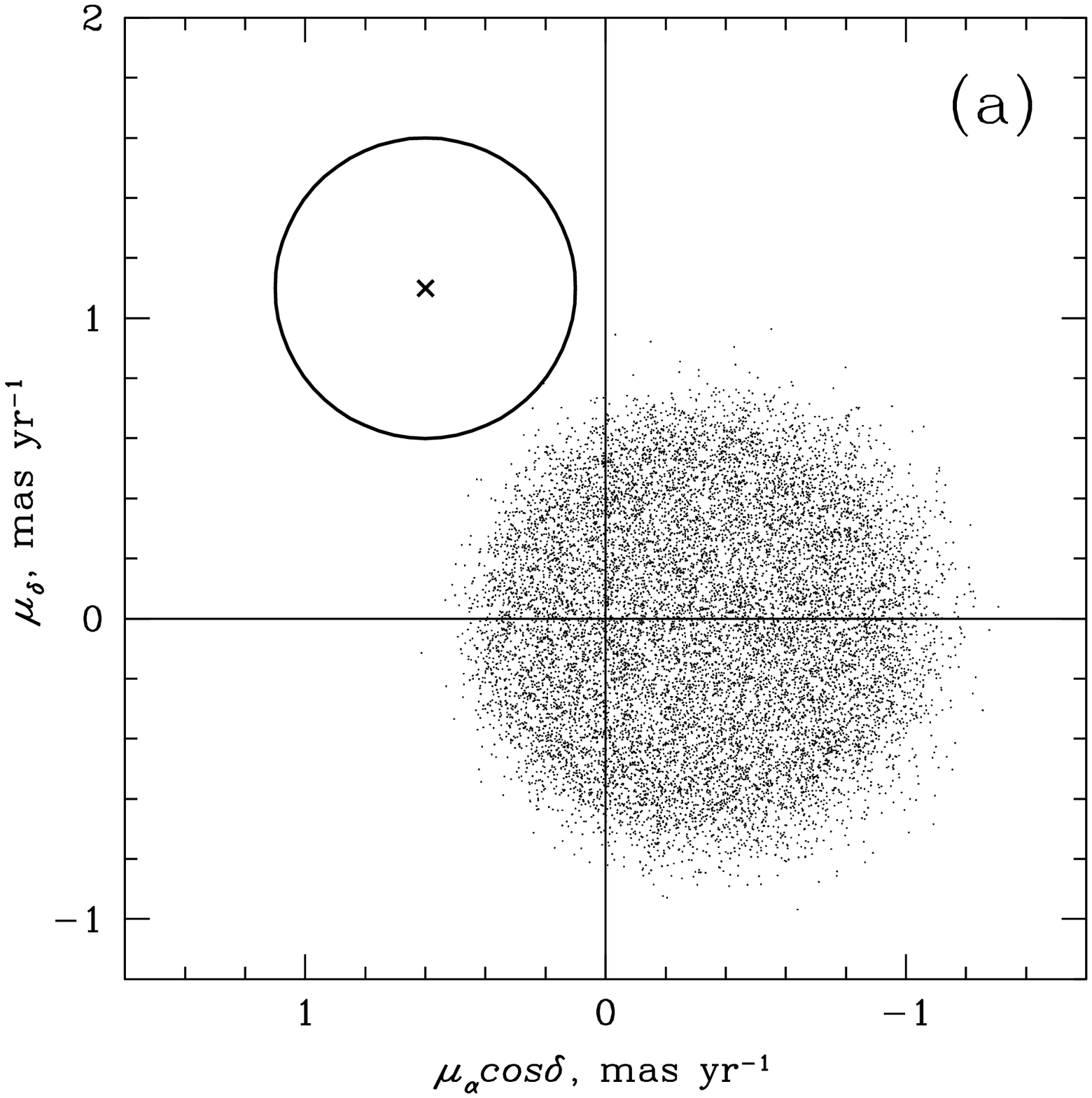}{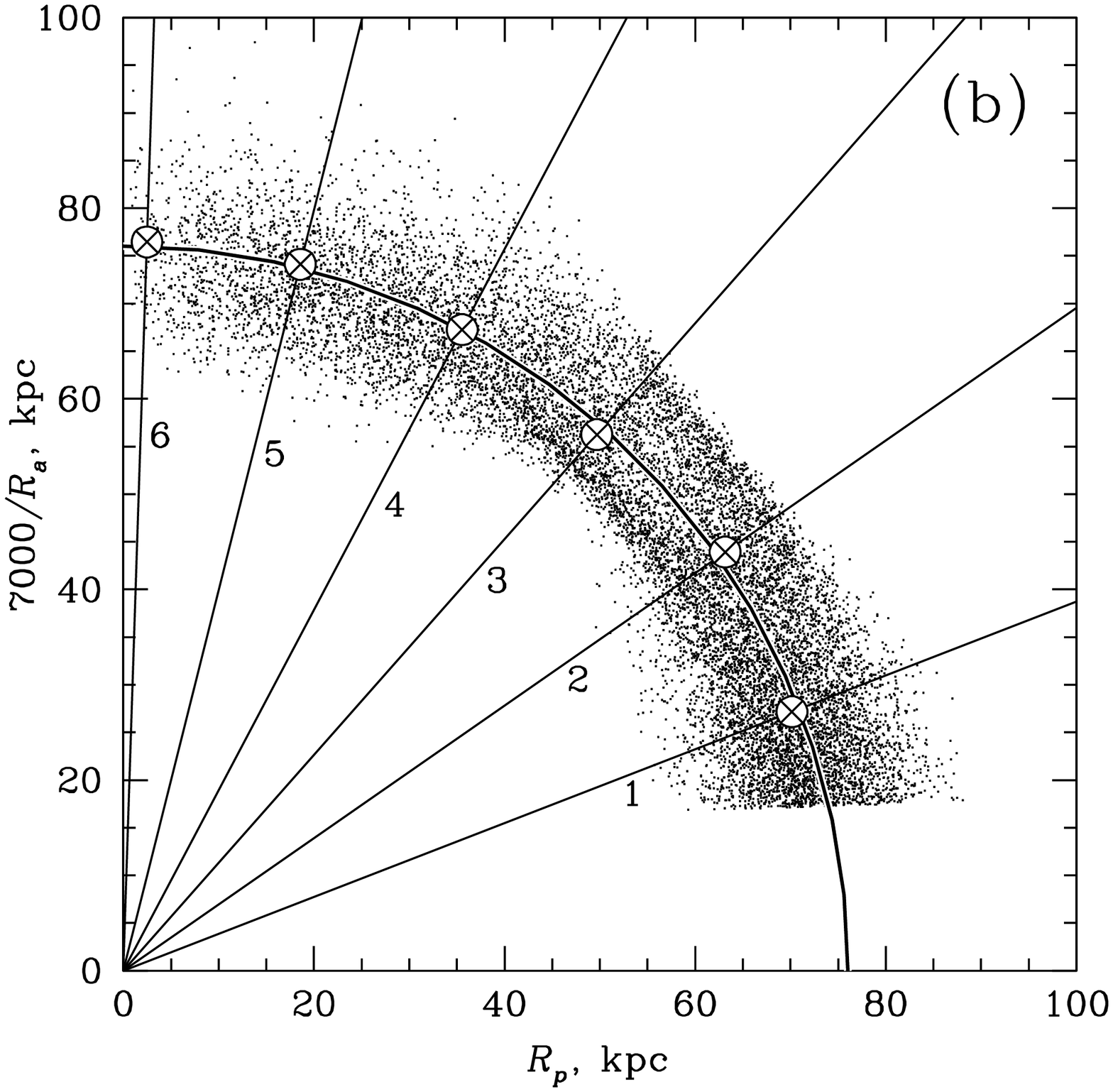}
\caption {(a) Proper motion vectors for Draco. The observational result of \citet{sch94} is shown as a cross
with the circle representing a 0.5~mas~yr$^{-1}$ one-sigma error bar. Dots
correspond to proper motion vectors which result in a bound orbit with the
period $P<8$~Gyr in the Milky Way potential.  (b) Draco's bound orbits with
$P<8$~Gyr in the ($R_p,7000/R_a$) coordinates (dots). In these coordinates, the
orbits follow well a circle with a radius 76~kpc (thick solid line). The cutoff
at $7000/R_a\sim 20$~kpc is due to the imposed cutoff in the orbital period
$P<8$~Gyr.  For six different values of the polar angle $\theta$ (numbered
radially divergent straight lines) we plot the averaged distances of the points
from the reference point (circles with crosses). As one can see, all the
averaged distances are very close to 76~kpc.
\label{fig5} }
\end{figure*}

\citet{sch94} published the only available measurement of the proper motion of Draco:
$\mu_\alpha\cos\delta=0.6\pm 0.5$~mas~yr$^{-1}$ and $\mu_\delta=1.1\pm
0.5$~mas~yr$^{-1}$.  (We adopted the larger value for the uncertainty from the
text of their paper.) When we started this project, we hoped that the
measurements of \citet{sch94} could be used to place useful constraints on
possible Draco's orbits in the Milky Way potential. Unfortunately,
this is not the case. This can be seen from Figure~\ref{fig5}a, which shows the proper motion measurements of
\citet{sch94} and the locus of the possible proper motion vectors resulting in a closed orbit
in the Milky Way potential given by equation~(\ref{pot}). (We assumed that the
halo density drops to zero beyond the virial radius $r_{\rm vir}$, and took into
account the uncertainties in the distance to Draco, $d=82\pm 6$~kpc, and its
line-of-sight velocity, $V_{\rm los}=-293\pm 2$~km~s$^{-1}$, which were taken
from \citealt{mat98}.) One can see that the observational results are virtually
inconsistent with Draco moving along a bound orbit around the Milky Way. More
quantitatively, the chance for a bound orbit with the radial period $P<8$~Gyr is
only 5.5\% (or 7.4\% for any period, which includes periods much longer than the
Hubble time). This is not surprising, as the proper motion measurements of \citet{sch94}
imply that Draco moves in the Galactic halo with a staggering speed of $610\pm
190$~km~s$^{-1}$ (one sigma error bars), with no realistic Milky Way model being
able to keep it gravitationally bound.

Given that Draco appears to be a pretty normal dwarf spheroidal galaxy and that
the dwarf spheroidals strongly concentrate toward the two large spirals in the
Local Group \citep*{mcb04}, the Milky Way and the M31 galaxy, it seems to be very unlikely that
this dwarf moves along an unbound orbit around our Galaxy, and just by chance
happened to be at the present small distance from the Galactic center.
We assume instead that Draco moves on a bound orbit, with $P\lesssim 8$~Gyr,
and that the proper motion measurements of \citet{sch94} are wrong. 
Proper motion measurements of the outer Galactic halo objects based on heterogeneous
collection of photographic plates spanning a few decades, such as those of
\citet{sch94}, are notoriously difficult to correct for all possible sources of systematic
errors. As another example, the proper motion measurement of the same authors
\citep{sch94} for Ursa Minor is more than one sigma away from each of the three more recent
results, including two derived with the help of the Hubble Space Telescope
\citep[their Fig.~16]{pia05}.

Fortunately, even in the case when the proper motion is not known, some
constraints can be placed on the orbital elements of Draco. We noticed that 
pericentric and apocentric distances for all the bound orbits from Figure~\ref{fig5}a 
are not completely independent, and follow closely
the solution of the equation

\begin{equation}
\label{rarp}
\left[R_p^2 + (7000/R_a)^2\right]^{1/2} = 76 \mbox{~kpc},
\end{equation}

\noindent which is a circle with the radius of 76~kpc in
the coordinates $x=R_p$, $y=7000/R_a$ (see Figure~\ref{fig5}b). (Here $R_p$ and
$R_a$ are in kpc.) As one can see in Figure~\ref{fig5}b, despite the
observational uncertainties in $d$ and $V_{\rm los}$ and the fact that the
proper motion in not known, the Draco's bound orbits stay in a relatively narrow
zone near the solution of equation~(\ref{rarp}). Neglecting the spread of the
points around the circle, one can assume then that bound Draco's orbits form a
one-dimensional family of models, with both $R_p$ and $R_a$ depending only on
the polar angle $\theta$, with $\tan\theta=y/x=7000/(R_p R_a)$.

\begin{figure*}
\plottwo{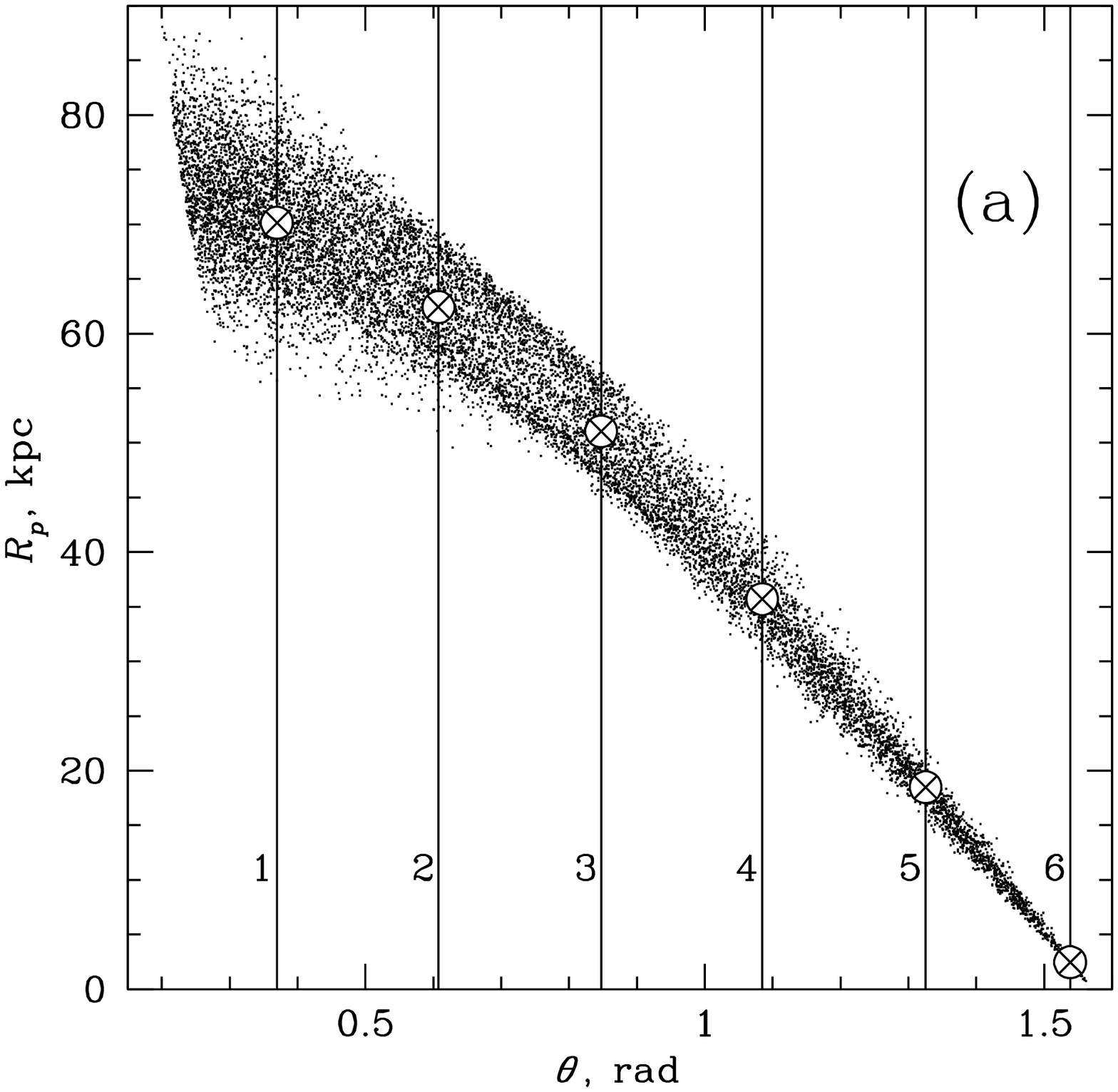}{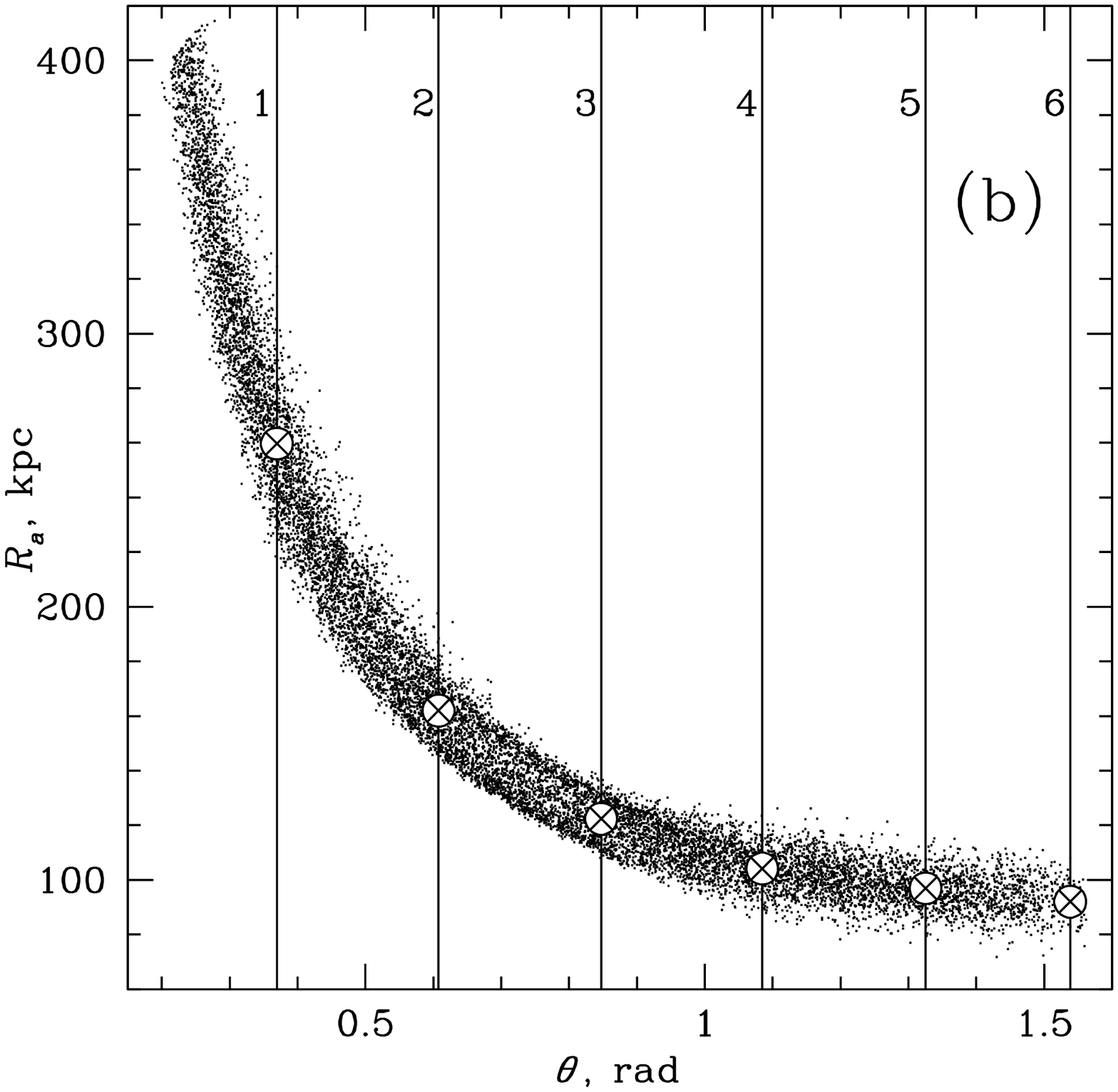}
\caption {Our choices for Draco's orbits. Dots show bound orbits with $P<8$~Gyr.
Vertical lines correspond to the six different values of the polar angle $\theta$ from
Figure~\ref{fig5}b.  Circles with crosses mark the averaged orbital parameter
(either pericentric distance $R_p$, panel a, or apocentric distance $R_a$, panel b) values
for different angles $\theta$.
\label{fig6} }
\end{figure*}

\begin{table}
\begin{center}
\caption{Draco's orbits\label{tab2}} 
\begin{tabular}{cccccccc}
\tableline
Orbit & $\theta$ & $R_p$ & $R_a$ & $R_a/R_p$ & $P$  & $V_{t,a}$ & $f_P$ \\
      &   rad    & kpc   &  kpc  &           & Gyr  & km~s$^{-1}$ &\\
\tableline
1     & 0.37     & 70.1  & 260   & 3.7       & 4.86 & 78.3  & 0.25\\
2     & 0.61     & 62.4  & 162   & 2.6       & 2.99 & 103.9 & 0.47\\
3     & 0.85     & 51.1  & 122   & 2.4       & 2.19 & 112.1 & 0.76\\
4     & 1.09     & 35.7  & 104   & 2.9       & 1.71 & 99.9  & 0.64\\
5     & 1.33     & 18.5  & 96.9  & 5.2       & 1.42 & 65.5  & 0.57\\
6     & 1.54     & 2.47  & 92.1  & 37        & 1.23 & 11.5  & 0.54\\
\tableline
\end{tabular}
\tablecomments{Here $V_{t,a}$ is the tangential velocity at $R=R_a$
and $f_P$ is the fraction of the time Draco spends at the distances
$R=70\dots 140$~kpc from the Galactic center, where 75\%
of Galactic dwarf spheroidals are currently located.
}
\end{center}
\end{table}

We chose six different values of the polar angle $\theta$ (shown as radially divergent
lines in Figure~\ref{fig5}b, and listed in Table~\ref{tab2}) to obtain a
sequence of bound Draco's orbits. We used plots shown in Figure~\ref{fig6} to
estimate the typical values of $R_p$ and $R_a$ corresponding to the particular
values of the angle $\theta$. In Table~\ref{tab2} we list the parameters of the
derived Draco's orbits. The range of covered orbital periods is $\sim 1-5$~Gyr.
All the orbits except for the last one have a pericenter at the distances where
the tidal disruption properties of our NFW halo are comparable to those of the
composite Milky Way model of \citet[see Figure~\ref{fig7}]{joh99}. The orbit 6
was designed to explore the extreme case of a virtually radial orbit. The
apocenter of the orbit with the longest period is at 260~kpc, which is at the
very edge of the virialized Galactic halo.  As one can see from
Table~\ref{tab2}, the derived sequence of orbits is not trivial, with the orbits
being more eccentric for longest and shortest periods, and becoming rounder for
intermediate periods of $\sim 2$~Gyr.

It is interesting to note that six out of eight, or 75\% of the ``classical'' Galactic dwarf
spheroidals (we exclude Saggitarius as being currently tidally disrupted) are
located in a rather narrow interval of Galactocentric distances $R=70\dots 140$~kpc
\citep{mat98}. This includes Draco, and excludes Leo~I and II. In Table~\ref{tab2}
we list for each orbit the fraction of time $f_P$ Draco spends in this interval
of Galactocentric distances. Statistically speaking, if Draco is a ``normal''
dwarf spheroidal, $f_P$ should be around 0.75. Our orbit 3 is in this sense the
likeliest orbit for Draco, and orbit 1 (and probably 2) are rather unlikely.

In Figure~\ref{fig3} we show as long-dashed lines the halos with the analytical
tidal radius $r_{\rm tid}$ equal to 0.85~kpc, which is the radius where the
density of Draco stars on the map of \citet{ode01} is two sigma above the noise
level. We calculated $r_{\rm tid}$ from

\begin{equation}
\frac{m(r_{\rm tid})}{r_{\rm tid}^3} = \left[ 2-\frac{R}{M(R)}\frac{\partial M}{\partial R}\right]
\frac{M(R)}{R^3}
\label{eq_rtid}
\end{equation}

\noindent \citep{hay03}, where $M(R)$ and
$m(r)$ are the enclosed mass for the Milky Way and the satellite,
respectively. Even at this large distance from the Draco's center \citet{ode01}
did not see any sign of Draco being tidally distorted by the Milky Way tidal
field. From Figure~\ref{fig3} one can see that all our best-fitting models
(solid circles) have tidal radii larger than 0.85~kpc (even for the worst
orbit with $R_p=2.5$~kpc).  One would naively conclude that the tidal forces are
not important for our Draco models.  But the reality is more complicated than
that. Numerical $N$-body simulations of the evolution of subhalos orbiting in
the host halo showed that removal of DM from the outskirts of the satellite
results in the expansion of the satellite, which reduces its average density and
exposes more DM to the action of the tidal field \citep{hay03,kaz04}.  Burkert halos
have lower average density than NFW halos, and as a result are even easier to
disrupt tidally \citep{mas05b}. Moreover, even well inside the tidal radius,
distribution of bound stars can be noticeably distorted by the tidal filed of
the Milky Way, which would be at odds with the Draco observations of
\citet{ode01}. To correctly describe the above effects, we had to resort to
high-resolution $N$-body simulations of the tidal disruption of our best-fitting
composite (DM $+$ stars) models from Table~\ref{tab1} in the static potential of
the Milky Way. This will be described in the following sections.

\subsection{Isolated Models}
\label{isol}

To run the $N$-body simulations of the tidal stripping of Draco in the static
gravitational potential of the Milky Way, we used the parallel tree-code Gadget-1.1
\citep*{spr01}. We generated equilibrium DM halos for our eight models 
(see Table~\ref{tab1}) using the prescription in \citet{mas05a}. The essence of
this approach is to use explicitly the distribution function (DF) to set up the
initial distribution of velocity vectors of DM particles (isotropy was assumed).
It was argued \citep{kaz04} that using DFs explicitly is far superior to
traditional local Maxwellian approximation for cuspy models such as NFW.  To
reduce the boundary effects, we truncate DM halos at a distance of two virial radii $r_{\rm
vir}$ from the center. This results in virtually no evolution for isolated
models within the virial extent of the halos after 10~Gyr. We chose the
gravitational softening lengths (separately for DM and stars) to be
commensurable with the average interparticle distance: $\varepsilon=0.77 r_h N^{-1/3}$
\citep{hay03}. Here $r_h$ is the half-mass radius of the system. For stellar particles,
$\varepsilon_*=8.6$~pc. For DM particles, $\varepsilon_{\rm DM}$ values are
listed in Table~\ref{tab1}.

To set up the initial distribution of stars inside the DM halo, we use the same
pseudo-Maxwellian approximation we used to measure the projected line-of-sight
velocity dispersion profiles in \S~\ref{st_model} (step 3). The only difference
is that now we use a larger number of stellar particles, $N_*=30,000$, to
reduce the Poisson noise in the observable properties of the models.

Our baseline number of DM particles is $N_{\rm DM}=10^5$. Test runs showed that
in the most massive models (N1,2 and B1,2) a much larger number of DM particles is
required to prevent the artificial evolution of the stellar cluster at the
center of the halo. We observed the central stellar density being significantly
reduced (by more than an order of magnitude in the worst cases) in the
low-resolution models. This effect does not depend on the number of stellar
particles, and becomes less severe for larger $N_{\rm DM}$ and/or
$\varepsilon_{\rm DM}$. As stars are only a trace population in our models, the
most obvious explanation for the above artifact is that stars get scattered from
the imperfections of the granulated gravitational potential of the DM halo.
Setting $N_{\rm DM}=10^6$ for the models N1 and B1, and $N_{\rm DM}=3.16\times
10^5$ for the models N2 and B2 resulted in acceptable level of the artificial
evolution of the surface brightness profile $\Sigma(r)$ of the stellar
cluster. In all our isolated models, the change in the central surface
brightness was negligible after 10~Gyr of evolution (with the only exception of
the model B1, where the change was $-0.4$~dex). The radius corresponding to the
outmost reliable Draco isodensity contour of \citet{ode01} with
$\Sigma=12,060$~M$_\odot$~kpc$^{-2}$ ($r=0.85$~kpc initially) increased by mere
$0.03-0.05$~dex.

Unfortunately, we observed significant evolution in the velocity anisotropy
profiles $\eta(r)$ in our isolated models, especially at the center of the halo.
All our best-fitting stellar models have a strong radial anisotropy at the
center (see Table~\ref{tab1}). In our $N$-body simulations with live DM halos
the stellar core becomes close to isotropic within the first Gyr, suggesting
that the particle-particle interactions are not the main culprit, as such
effects would take Gyrs to manifest themselves. This became more apparent after
we ran simulations for all our models from Table~\ref{tab1} with $N_*=30,000$
stellar particles and {\it static} DM potential. In this setup, close encounters
between particles are extremely rare due to very low stellar density. In the
static models we see the same effect (of slightly smaller magnitude) as in the
runs with live DM halos: central radial anisotropy reduces almost to zero after
a few crossing times. We do not know the exact reason for the above effect. The
possible explanations are (1) the (unknown) DF corresponding to our choice of anisotropy,
stellar density, and gravitational potential profiles is okay (positive everywhere), but the
pseudo-Maxwellian approximation we use to set up the stellar velocities is not
good for systems with strong radial anisotropy at the center, and/or (2) the DF
is not physical (i.e. is negative) at the center. We also want to point out that
a real stellar system cannot have a radial anisotropy
all the way to its center, as the radial velocity dispersion diverges in such
case when $r\rightarrow 0$. To demonstrate this, we write down the solution of
the Jeans equation~(\ref{Jeans2}) in the case of constant anisotropy
\citep{lok01}

\begin{equation}
\label{const}
\sigma_r^2 = \frac{1}{r^{\zeta}\rho_*}\int\limits_r^\infty r'^\zeta \rho_* \frac{d\Phi}{dr} dr',
\end{equation}

\noindent where the constant $\zeta\equiv 2\beta=4\eta/(1+\eta)$ is positive for the case of
radial anisotropy. The integral in equation~(\ref{const}) is non-zero for
$r=0$, resulting in divergent $\sigma_r$ at the center of the system.
Obviously, in a real object radial anisotropy should break down at some radius,
changing into isotropy or tangential anisotropy. It is also hard to expect
radially divergent velocity dispersion in a stellar system with a flat core,
especially in the case of a flat-cored (Burkert) DM halo.

\begin{figure}
\plotone{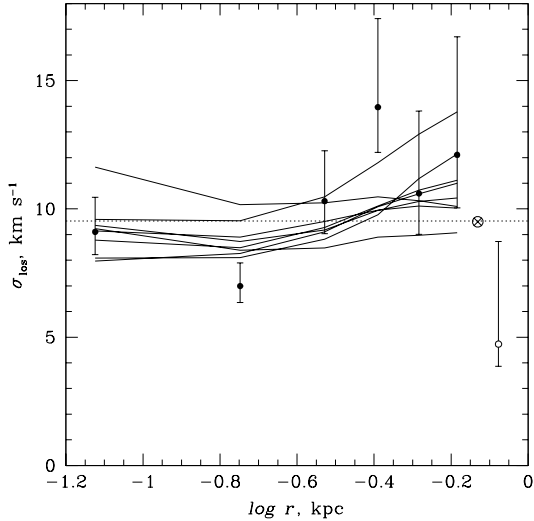}
\caption {Line-of-sight stellar velocity dispersion profiles for Draco for the
models evolved in a static DM halo potential. The notations are the same as in
Figure~\ref{fig4}.
\label{fig8} }
\end{figure}

We want to emphasize that even though we cannot guarantee that the stellar models
in Table~\ref{tab1} are physical (especially at the center), in our simulations
they quickly relax to a stable configuration, with the surface brightness
profile virtually identical to the observed one, and the line-of-sight velocity
dispersion profile $\sigma_{\rm los}(r)$ still consistent with the
observations. In Figure~\ref{fig8} we plot $\sigma_{\rm los}(r)$ profiles for
our stellar models in a static DM potential after 10~Gyr of evolution integrated
over the same six radial bins as the observational data of \citet{wil04}. As one
can see, the profiles for evolved stellar clusters are a reasonably good match
to the observations. (The $\chi^2$ measure becomes a factor of two larger in the
evolved models.)

We also used the isolated runs to estimate the accumulated total energy errors
in our models. In the case of live DM halos, the total energy for DM $+$ stars
is conserved to better than $\sim 0.1$\% (typically $\sim 0.05$\%) after 10~Gyr
of evolution. To estimate the total energy errors for stars only, we measured
the difference in stellar total energy at $t\sim 3$~Gyr (when the cluster has
reached a steady state configuration) and at $t=10$~Gyr in our static DM halo
simulations. The difference was again $\lesssim 0.1$\%.

\subsection{Tidal Stripping Simulations}

We simulated evolution of the eight stars$+$DM models of Draco from
Table~\ref{tab1} orbiting for 10~Gyr in the static spherically symmetric
potential of the Milky Way given by equation~(\ref{pot}). To run the
simulations, we used the parallel version of the multi-stepping tree code Gadget
\citep{spr01}. The number of stellar and DM particles and the corresponding gravitational 
softening lengths were the same as in the isolated models described in
\S~\ref{isol}. Each Draco model was simulated for six different orbits from Table~\ref{tab2}.
Initially, Draco was located at the apocenter of its orbit.
Altogether we made 48 tidal stripping simulations. We refer to the models by
both the model number from Table~\ref{tab1} and the orbit number from
Table~\ref{tab2}, e.g. model N1-5. Most of the results we give below are
for the latest moment of time when the dwarf was located at the current distance
of Draco from the Galactic center of $\sim 82$~kpc (we do not make distinction
between the dwarf moving inward or outward), which took place $7.4-10$~Gyr from
the beginning of the simulations ($9.4-10$~Gyr for the orbits $3-6$).

\begin{figure*}
\plottwo{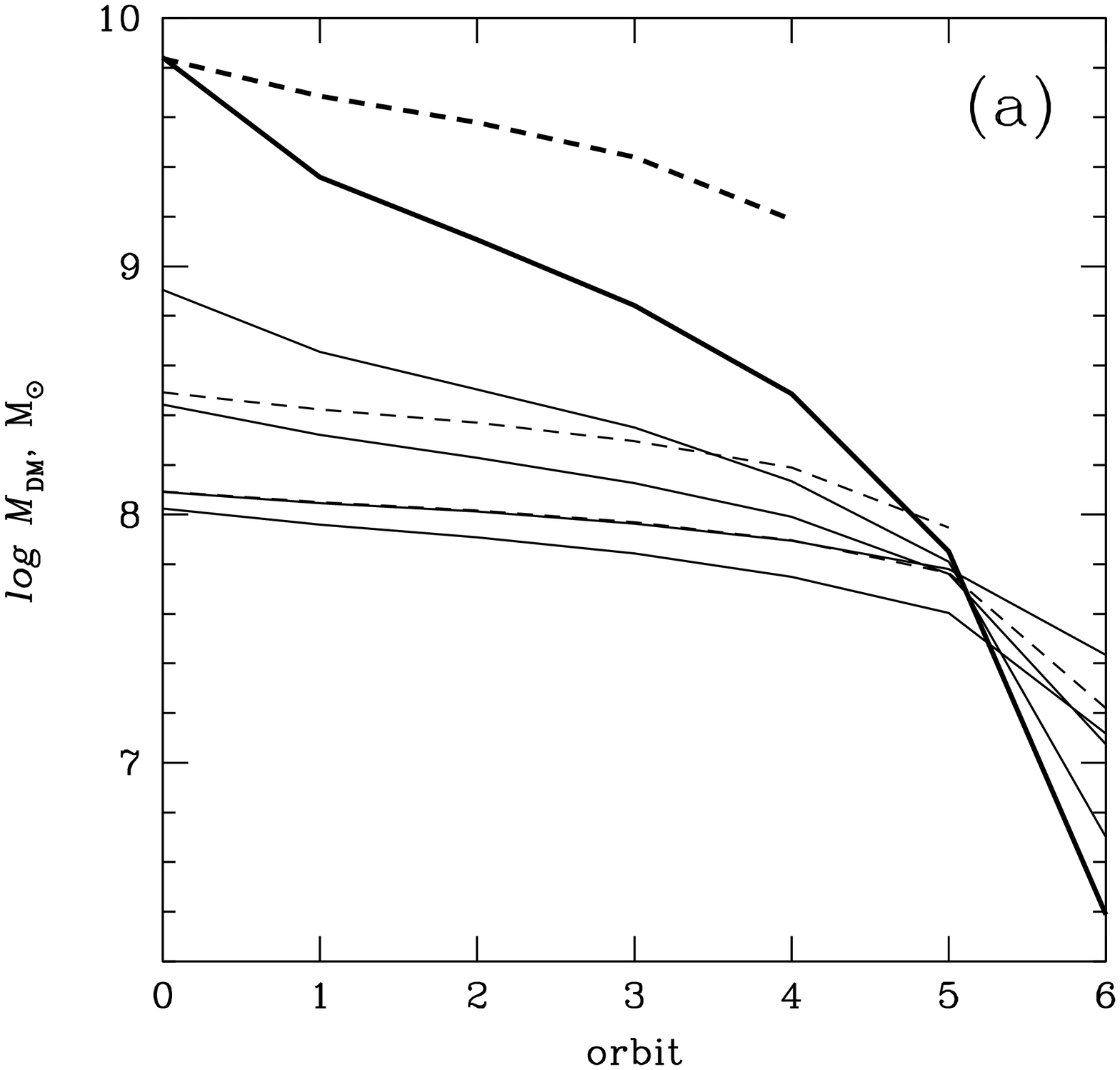}{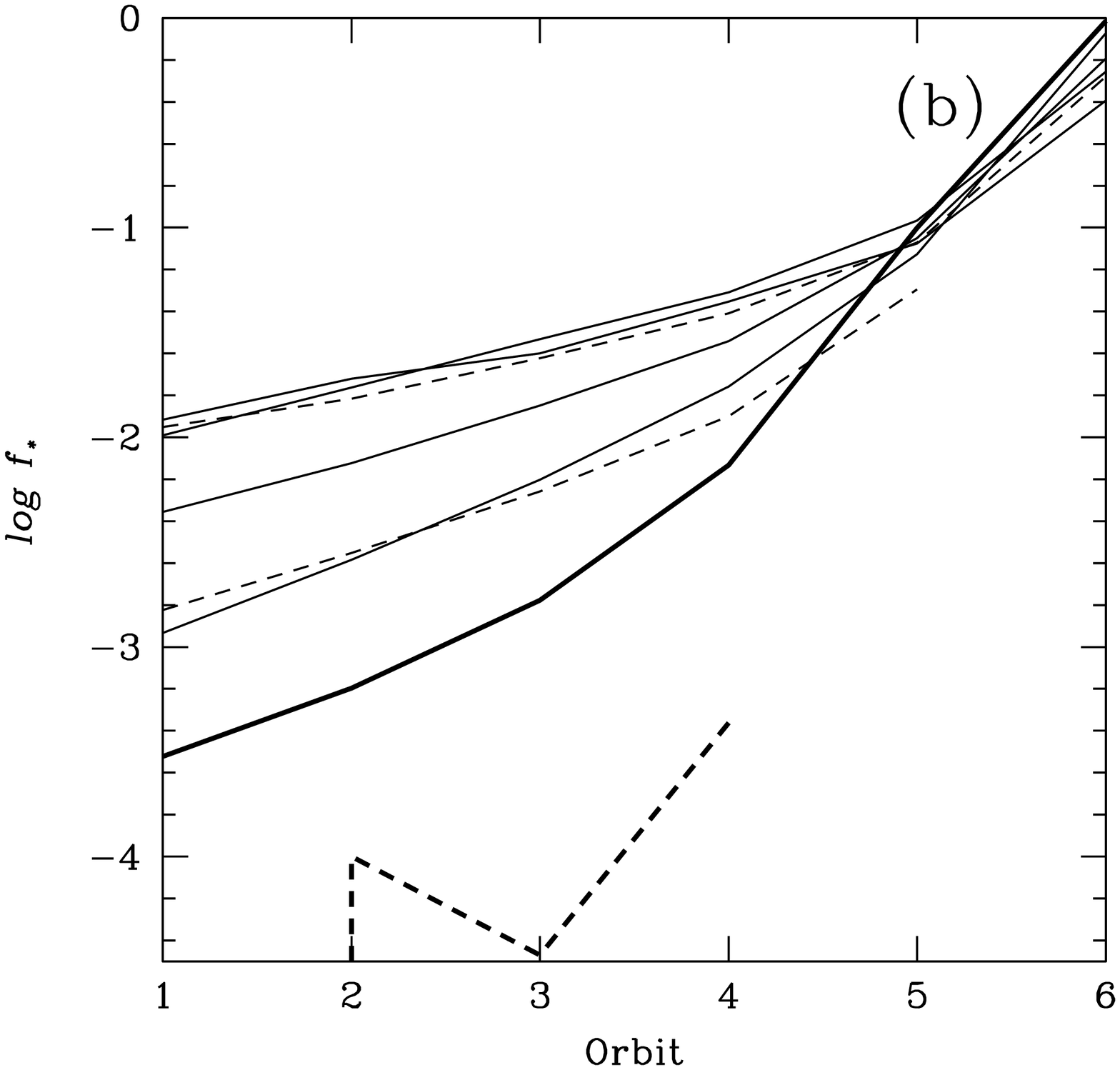}
\plottwo{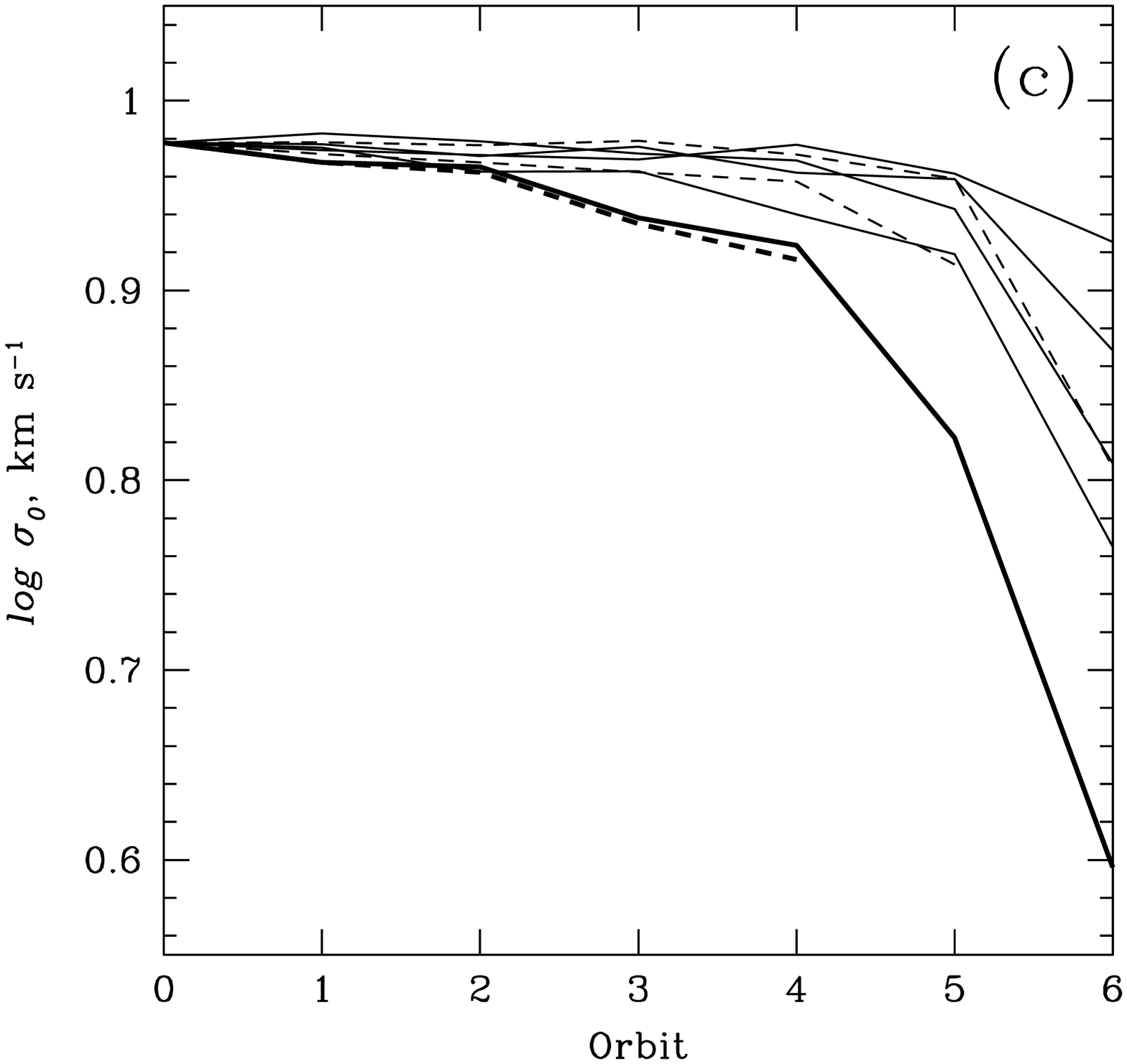}{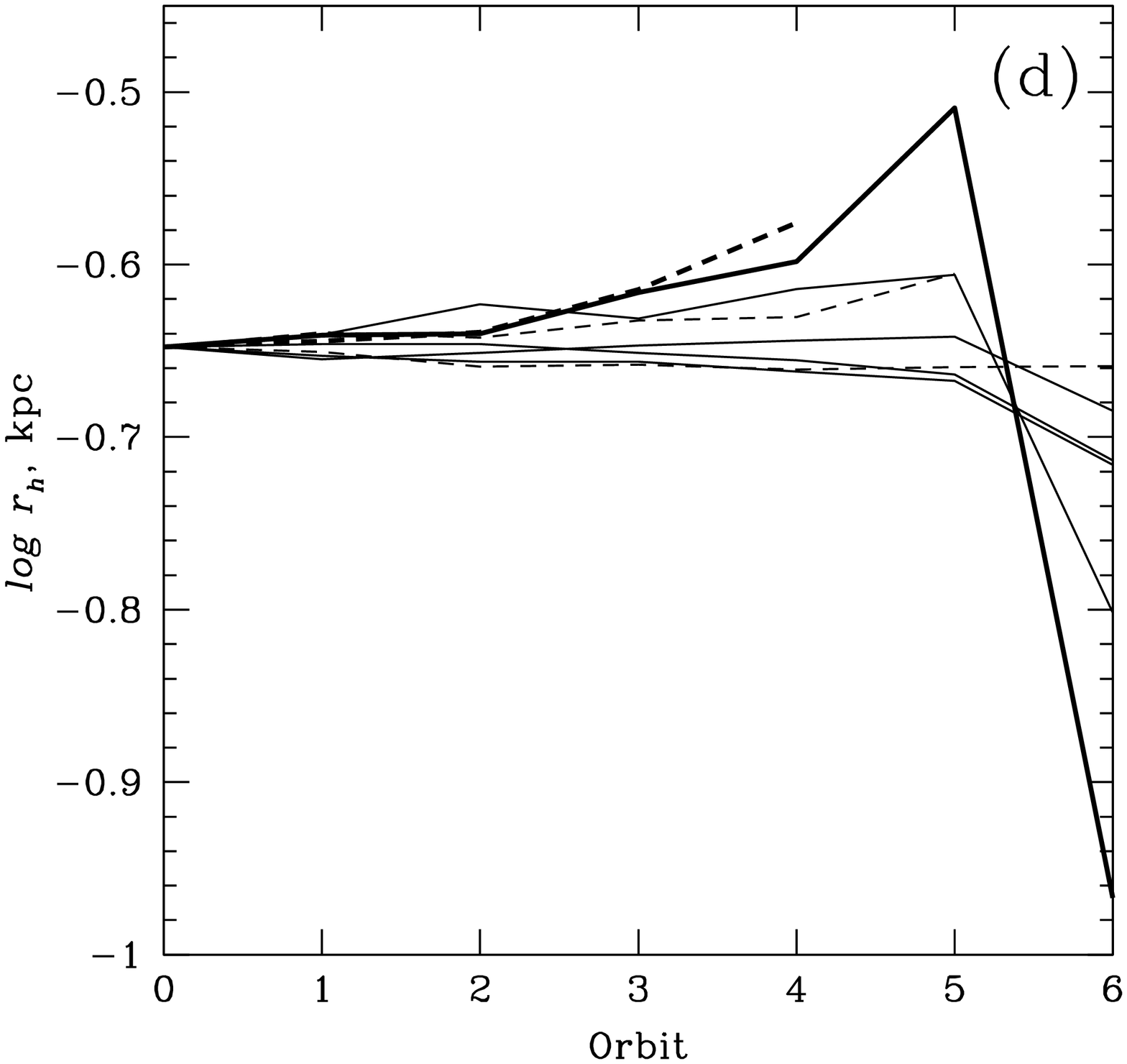}
\caption {Draco models' parameters near the end of the simulations
(when the dwarf was $\sim 82$~kpc away from the Galactic center) as a function
of orbit. Solid/dashed lines correspond to models with NFW/Burkert DM halo
profiles, respectively. Thick lines correspond to the most massive halos (models
N1 and B1). (a) Gravitationally bound DM mass $M_{\rm DM}$. (b) Fraction of
stars having become unbound $f_*$. (c) Central line-of-sight stellar velocity
dispersion $\sigma_0$. (d) Projected half-light radius for the bound
stellar cluster $r_h$. To facilitate the comparison of different models, both
$\sigma_0$ and $r_h$ are normalized to the same value for the orbit 0
(corresponding to isolated models).
\label{fig9} }
\end{figure*}

All our models experienced tidal stripping of different degree. As
Figure~\ref{fig9}a shows, the DM mass of the gravitationally bound remnant is
between 90\% (model N5-1) and 0.1\% (model N1-6) of the original mass at the end
of the simulations. Three of our models became completely gravitationally
unbound within 10~Gyr: B1-6 (after 3.2~Gyr and 3 pericenter passages), B2-6
(after 7.8~Gyr and 6 pericenter passages), and B1-5 (after 9.4~Gyr and 7
pericenter passages). In the latter case, the dwarf is still bound when it is
located at $\sim 82$~kpc at the end of the simulations (when we compare the
results of the simulations with the observed properties of Draco) -- but only
barely so.  It is not surprising that all the unbound models have Burkert
halos. Indeed, for given mass and scaling radius these halos have lower averaged
density than NFW halos in the central area. If truncated
instantaneously at a certain radius, the total energy of the remnant becomes
positive for Burkert halos at a radius $\sim 2.1$ times larger than for
NFW halos \citep{mas05b}. 

Another result which can be explained is that the most massive halos are easier
to strip and disrupt tidally than the less massive ones. All our models (both
massive and of lower mass) have a comparable DM density within the observed
extent of Draco (because of the virial theorem), so in the point mass
approximation they should be equally susceptible to tidal forces.  However, the
point mass approximation (used to derive equation~[\ref{eq_rtid}]) breaks down
for our most massive halos, as their size becomes comparable to the pericentric
distance, so the strongly non-linear components of the tidal force become
important.  Not surprisingly, our most massive halos on the orbit 6 were either
totally disrupted (model B1-6) or lost 99.9\% of its original mass (model N1-6)
after 10~Gyr of evolution.

In all of our models a fraction of stars has been tidally stripped by the end of
the simulations -- even for the orbit 1. In two cases (models B1-6 and B2-6),
stars have become completely unbound by the time the dwarf was passing at the
distance of $\sim 82$~kpc from the Galactic center for the last time. In other
cases, the fraction of escaped stars was between $\sim 10^{-4}$ for the models
B1-2,3,4 and 96.6\% for the model N1-6 (see Figure~\ref{fig9}b).

When analyzing the global properties of the stellar cluster at the end of the
simulations, the most obvious result is the fact that the more disruptive the
orbit is, the more the system is affected by the tidal shocks experienced near
the pericenter of the orbit. (Orbits with a larger number from Table~\ref{tab2}
are more disruptive for two reasons: they have smaller pericentric distance, and
they have shorter orbital period, so the number of pericentric passages in
10~Gyr is larger.) Dwarfs are puffed up by the tidal shocks, with the central
line-of-sight velocity dispersion $\sigma_0$ becoming smaller
(Figure~\ref{fig9}c), central surface brightness decreasing (not shown, but is
qualitatively similar to $\sigma_0$ behavior), and the projected half-light
radius becoming larger (except for the extreme case of the orbit 6 when
both tidal shocking and tidal stripping are important, see Figure~\ref{fig9}d).

Our models are highly idealized when it comes to long-term evolution of stellar
tidal streams. On our most disruptive orbits 5 and 6, a significant fraction of
stars becomes unbound over the course of the tidal evolution, with the most of
the tidally stripped stars following very closely the almost radial orbit of the
dwarf. From the Sun location, many of these stars project on a rather small area
in the sky inside or around the apparent location of the dwarf. This can be
quite unphysical, as such cold tidal streams are not expected to survive for
many gigayears in the Milky Way halo due to its triaxiality and clumpiness
(as predicted by cosmology) and due to interaction with baryonic
structures in the Galaxy (stellar disk with spiral arms, stellar bar, giant
molecular clouds). To circumvent this difficulty, in this section we discuss all
the observable model properties for two extreme cases: (a) all stars are taken
into account, and (b) only the stars (both bound and unbound) located within the
spatial distance of 5~kpc from the center of the dwarf are considered. In the
latter case, the spatial truncation of the tidal stream ensures that only the
most recently stripped stars are used for calculating the observable properties
of the models.

\begin{figure*}
\plottwo{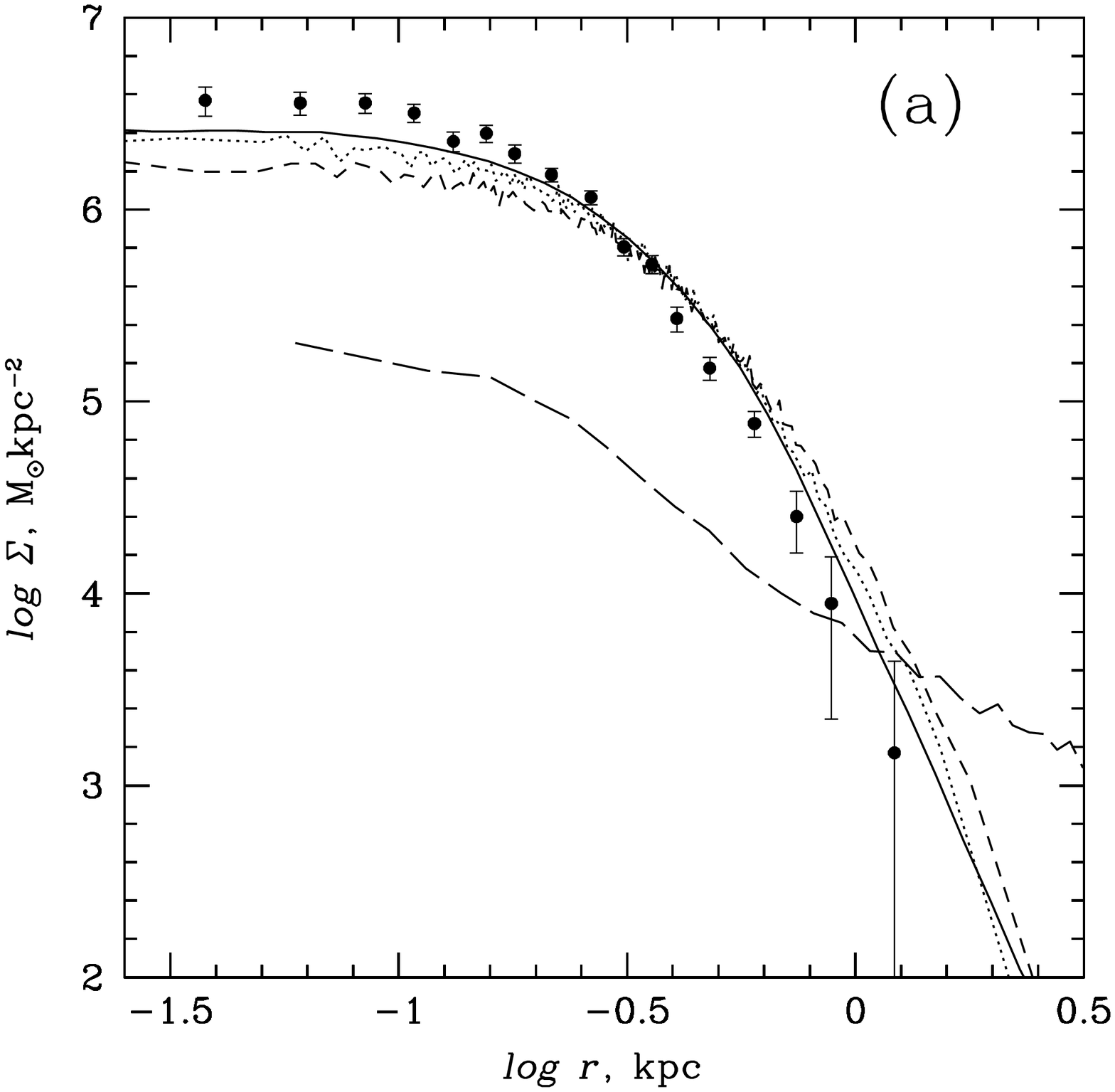}{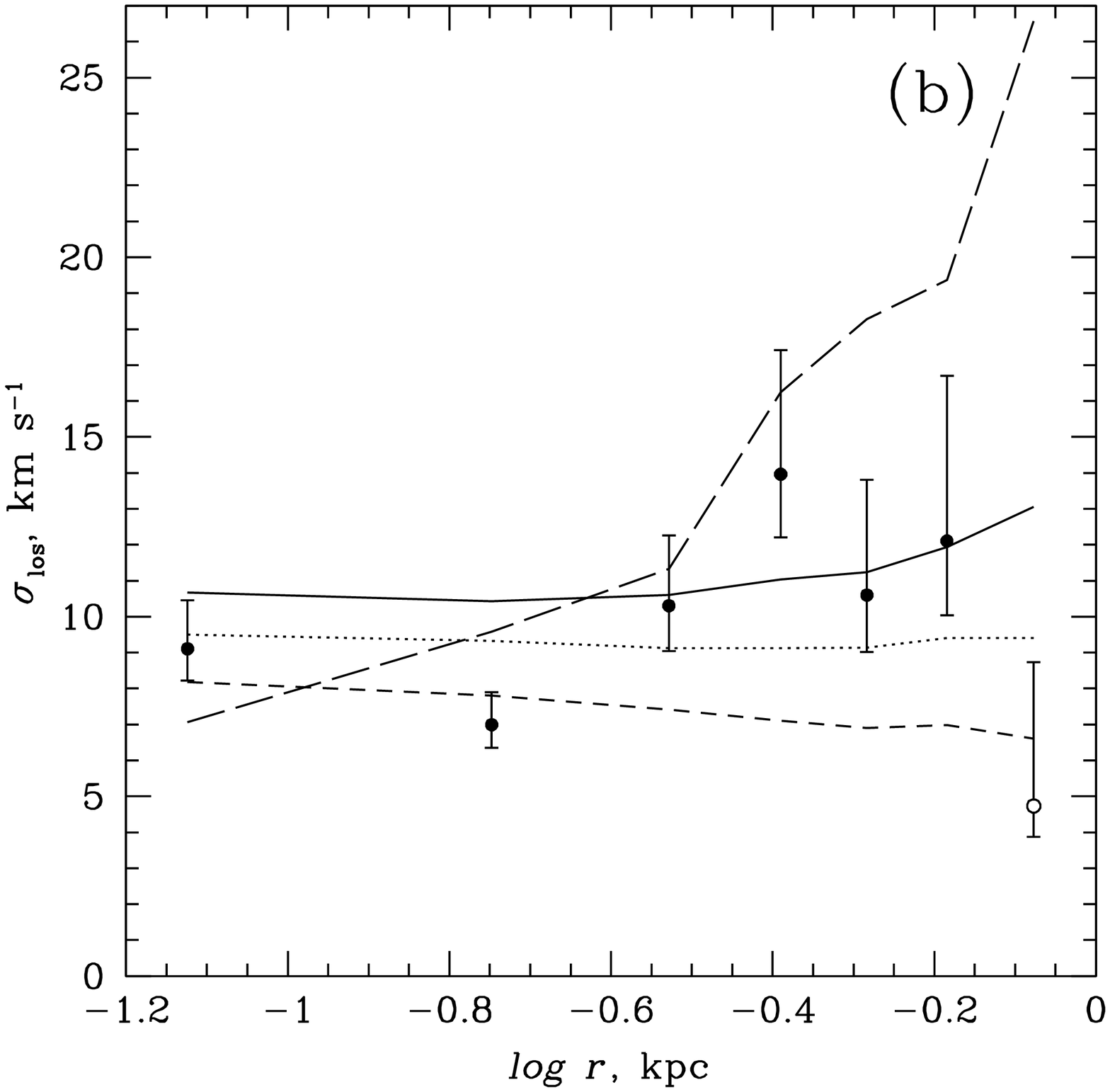}
\caption {Observable properties of the model N1 for a few different orbits near the end of the
simulations (when the dwarf was $\sim 82$~kpc away from the Galactic
center). Only stars (both bound and unbound) located within the spatial distance
of 5~kpc from the dwarf's center are considered.  Solid, dotted, short-dashed,
and long-dashed lines correspond to orbits 1, 4, 5, 6, respectively.  The
observer is assumed to be located at the Galactic center.  (a) Surface
brightness profile $\Sigma(r)$. We also show the observed Draco profile of
\citet[their sample S2]{ode01}. (b) Stellar line-of-sight velocity dispersion
profile $\sigma_{\rm los}(r)$. We also show the observed Draco velocities of
\citet{wil04}.
\label{fig11} }
\end{figure*}

On a more detailed level, the impact of both tidal shocks and tidal stripping
and heating can be seen in Figure~\ref{fig11}. Here we show the surface
brightness profiles (panel a) and line-of-sight velocity dispersion profiles
(panel b) for the models N1-1,4,5,6. The most obvious effect in
Figure~\ref{fig11} is the global decrease of $\sigma_{\rm los}$ for orbits with
smaller pericentric distance (excluding the orbit 6), accompanied by a small
decrease in the central surface brightness and a slight radial expansion of the
system. No obvious tidal truncation and tidal heating is observed in the
outskirts in the cluster.  These results were obtained for the stars located
within the spatial distance of 5~kpc from the center of the dwarf, but in the
case when all stars are included the profiles are practically the same. The
orbit 6 is a completely different case: one can see the $\sigma_{\rm los}$ being
dramatically inflated in the outskirts of the dwarf due to superposition of
tidally removed stars on the dwarf (Figure~\ref{fig11}b). This effect becomes
even more dramatic when we include all the stars, with the line-of-sight
velocity dispersion reaching $50-70$~km~s$^{-1}$ in the outermost observed
bin. As we discussed in the previous paragraph, it is quite unlikely that old
tidal streams can stay so well collimated for many gigayears to produce the
above effect. But even for the conservative case of considering only freshly
stripped stars the steeply growing $\sigma_{\rm los}$ profile for the model N1-6
is grossly inconsistent with the observed profiles of Draco and other dwarf
spheroidals, where $\sigma_{\rm los}$ is either not changing or decreasing at
large distances from the center.

The change in the surface brightness profile for the model N1-6 is also quite
dramatic (Figure~\ref{fig11}a, long-dashed line). The outer $\Sigma(r)$ slope becomes very shallow
which is inconsistent with the observed profiles for Galactic dSphs. The slope
becomes even more shallow when we include stars beyond the spatial distance of
5~kpc from the center of the dwarf. Similar behavior (in terms of shallow outer
$\Sigma$ profiles and inflated $\sigma_{\rm los}$ in the outskirts of the
dwarf) is also observed in other models with the orbit 6 (both bound and
unbound). The possible exceptions are the models N4-6 and N5-6, which
are consistent with the observations of Draco when we consider only freshly
stripped stars. We conclude that it is unlikely that Draco and other Galactic
dSphs have experienced tidal interactions as dramatic as our models on the
orbit 6.

To facilitate the comparison of our models with the observed stellar isodensity
contours of Draco of \citet{ode01} we designed the following projection
algorithm.  (a) The frame of reference is rotated to place the center of the
dwarf on the negative side of the axis $Z$, with the axis $X$ located in the
orbital plane of the dwarf and pointing in the direction of the orbital motion.
(b) As the direction of the proper motion of Draco is not known (see discussion
in \S~\ref{Orbits}) and the current angle between the vector connecting Draco
with the center of the Galaxy and the vector connecting Draco with Sun is
$\varphi=5\fdg 7$, we consider three extreme cases of the possible vantage point
location which should encompass the whole range of projected model appearances:
view from the Galactic center ($\varphi=0\degr$), view from a point in the
orbital plane of Draco located at 82~kpc from the dwarf with $\varphi=5\fdg 7$,
and view from a point in the plane which is perpendicular to the orbital plane
of Draco at 82~kpc and with $\varphi=5\fdg 7$. We found that due to the fact
that for Draco the angle $\varphi$ is small (which is also the case for other
Galactic dSphs with the exception of Sagittarius), the observable properties of
our models ($\Sigma$ and $\sigma_{\rm los}$ profiles and surface brightness
maps) do not depend noticeably on the vantage point we choose -- especially for
the case when we only include freshly stripped stars. (c) We exclude stars with
$z>-8.5$~kpc to avoid contamination of our maps with local tidal stream
overdensities which would be discarded by observers as local Galactic stars.
(d) We perform prospective projection of the stellar particles smoothed with a
Gaussian beam which has a fixed physical size (either 0.15 or 0.5~kpc) and
brightness inversely proportional to the square of the distance of the particle from the
observer. This procedure makes the ``surface brightness'' of individual
particles invariant of the distance from the observer, which is appropriate for
spatially resolved clumps of the tidal stream.

\begin{figure*}
\plottwo{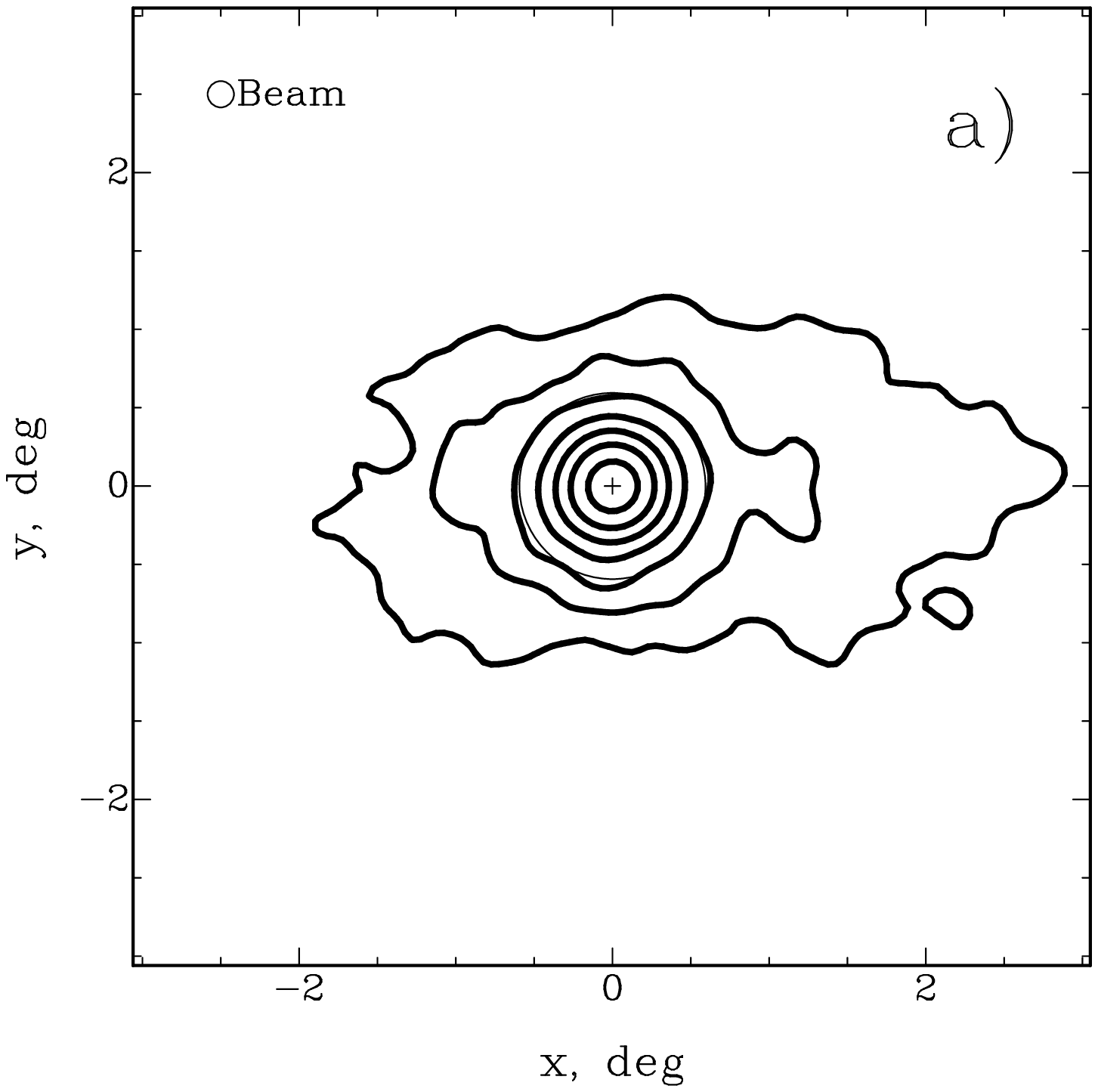}{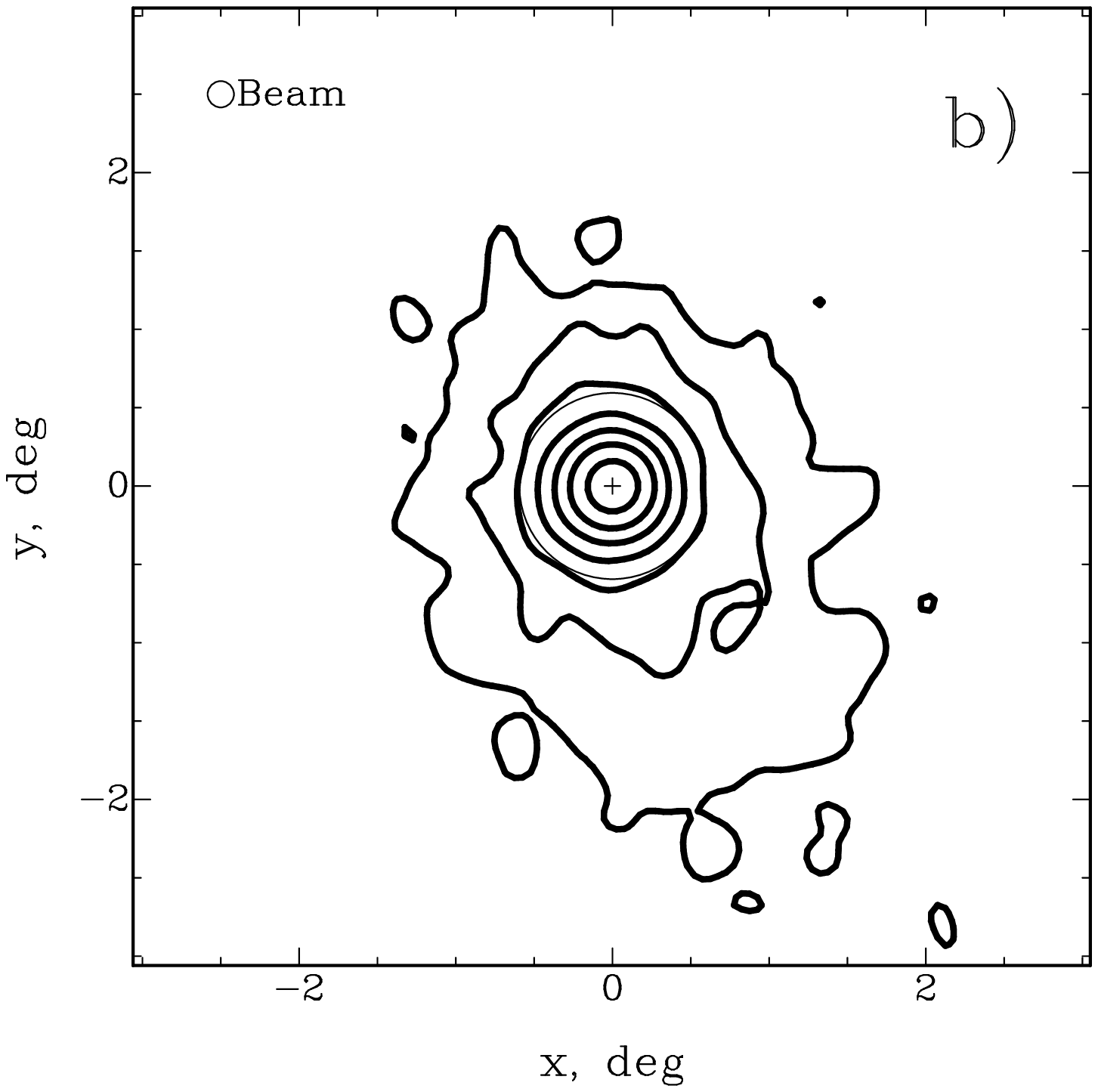}
\caption {Stellar surface brightness maps for the model N2-6 near the end of the
simulations (when the dwarf was $\sim 82$~kpc away from the Galactic
center). The contour levels are $(8,16,32,\dots,512)\times
10^3$~M$_\odot$~deg$^2$ (calculated for the nominal distance of Draco of
82~kpc). Cross marks the center of the dwarf. Thin line circle has a radius of
0.85~kpc -- the radius of the Draco two-sigma surface brightness contour of
\citet{ode01}. In the upper left corner we show the size of the Gaussian beam
used to make the maps. (a) The observer is in the plane of the Draco's
orbit. (b) The observer is in the plane perpendicular to the Draco's orbit.
\label{fig12} }
\end{figure*}

The most interesting result obtained from the analysis of the surface brightness
maps is the lack of the classical S-shaped tidal tails in our models. In the
case of significant tidal stripping (Figure~\ref{fig12}, two different vantage
points are shown) the stellar isodensity contours change from being spherical
near the center of the dwarf to being increasingly more elliptical and often
off-centered at larger distances. In the cases with less severe stripping, outer
contour boxiness is observed in some of the models. For orbits 1-3 the surface
density of the tidally removed stars is so low that it is hard to draw any
conclusion as to the shape of the tidally distorted isodensity contours (except
for the fact that the galaxy is observed against the background and/or
foreground of a few degrees wide belt of tidally stripped stars).  The
explanation for the above effect is in the facts that the tidal stripping is
significant only for very eccentric (almost radial) Draco orbits with
$R_p\lesssim 20$~kpc, and that currently Draco is located $\gtrsim 10$~kpc away
from the apocenter of its orbit (see Table~\ref{tab2}). Under these
circumstances, line of sight practically coincides with the direction of the
tidal elongation of Draco, with both tidal tails seen edge-on.  This is an
interesting result, as it suggests that the lack of S-shaped isodensity contours
in the outskirts of the Galactic dSphs cannot be used to support a claim that the dwarf
has not experienced significant tidal stripping in the gravitational potential
of the Milky Way. It appears that the presence (absence) of tidally heated stars
in the outskirts of dSphs is much more sensitive indicator of tidal stripping
being significant (not significant).

\begin{figure*}
\plottwo{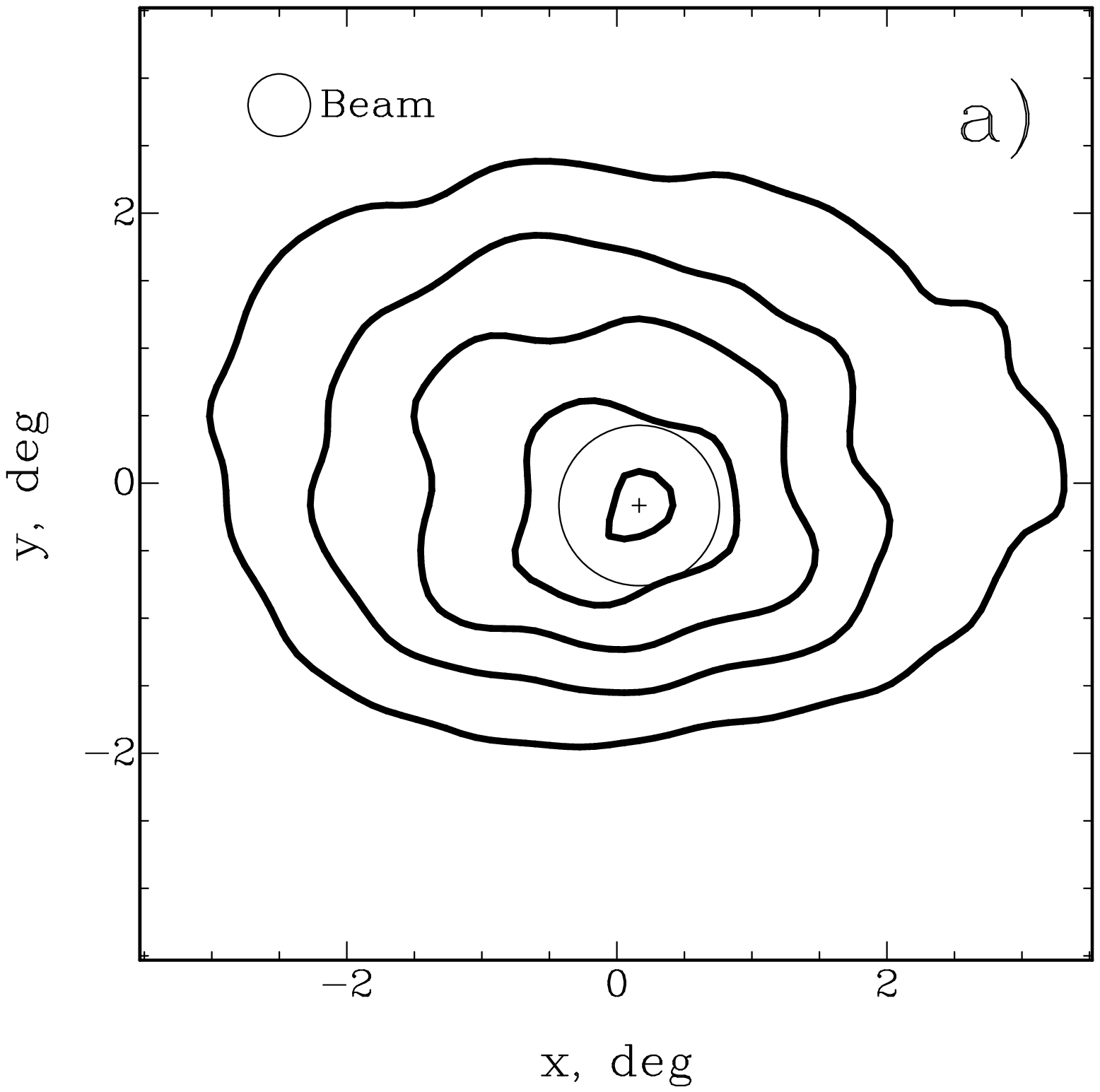}{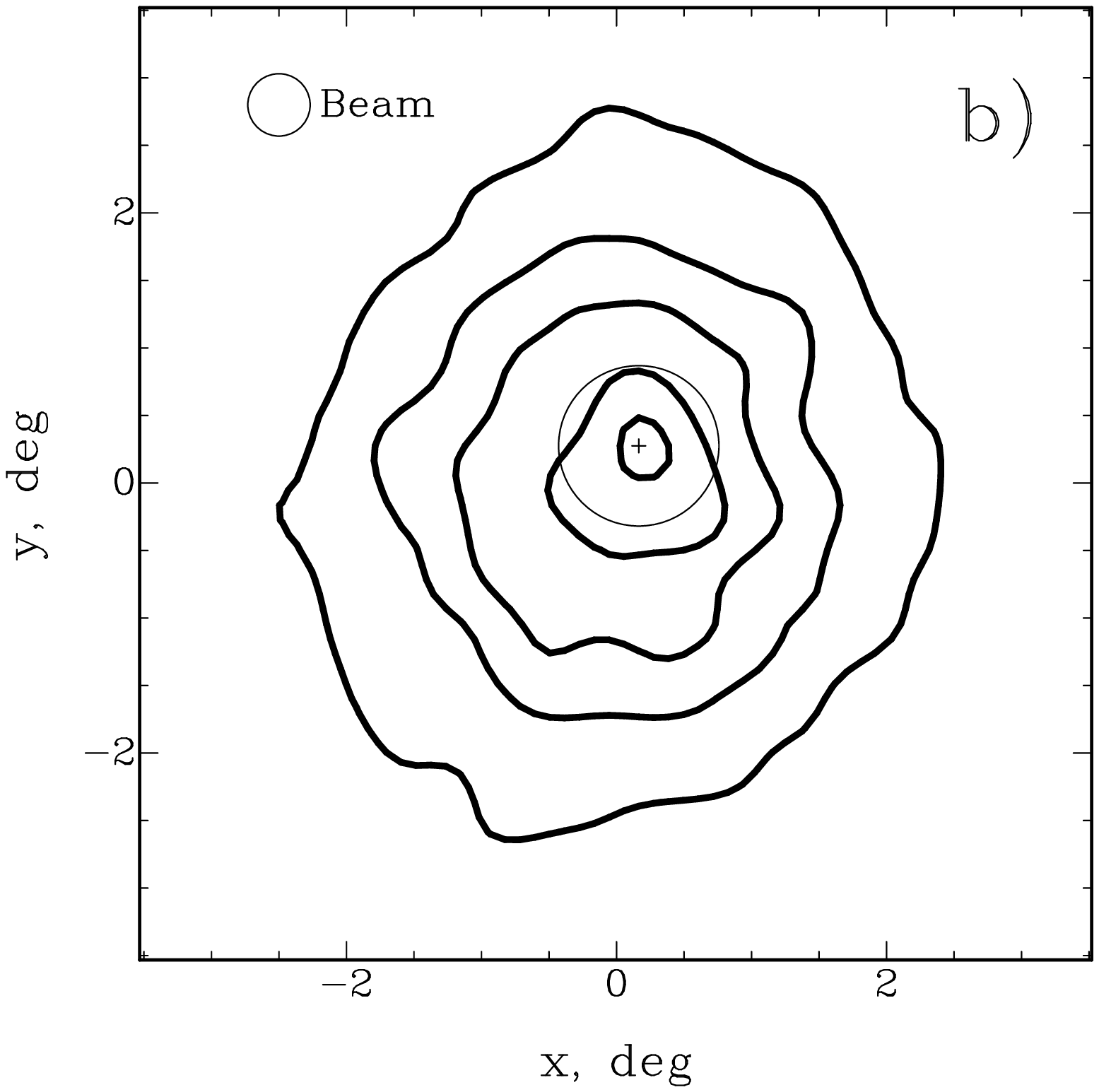}
\caption {Surface brightness maps for the two unbound models at the end of the simulations
(when the dwarf is located at $R\sim 82$~kpc). Only freshly stripped stars
(within a radius of 5~kpc from the dwarf) are included. The observer is located
at the center of the Galaxy. Cross marks the center of the dwarf. In the upper
left corner we show the size of the Gaussian beam used to make the maps. (a)
Model B1-6. Contours are $(4\dots 19)\times 10^3$~M$_\odot$~deg$^{-2}$.  (b)
Model B2-6. Contours are $(1\dots 8.6)\times 10^3$~M$_\odot$~deg$^{-2}$.
\label{fig15} }
\end{figure*}

We measured for all our models the critical surface brightness $\Sigma_c$ when
the outer isodensity contours become noticeably distorted due to tidal effects.
The two models which became gravitationally unbound by the end of the
simulations (B1-6 and B2-6) show very distorted contours all the way to the
center of the dwarf (see Figure~\ref{fig15}). They also have dramatically inflated line-of-sight stellar
velocity dispersion profiles (up to $50-100$~km~s$^{-1}$), and very shallow
outer surface brightness slopes. All of the above makes these two models (and
most probably any other unbound model for a dSph) completely inconsistent with
the observed properties of the Galactic dSphs. Among the bound models, the one
which has experienced the most dramatic tidal stripping (model N1-6) is also
the one which has the largest value for $\Sigma_c$: $\sim 4\times
10^4$~M$_\odot$~deg$^{-2}$, which would correspond to $\sim 3$ sigma isodensity
contour of \citet{ode01}. In all other cases, the value of $\Sigma_c$ is
significantly lower: $(7.4-17)\times 10^3$~M$_\odot$~deg$^{-2}$ for the models
on the orbit 6, and less than $10^3$~M$_\odot$~deg$^{-2}$ for other orbits. For
the orbits 1, 2, and 3 $\Sigma_c$ was too small to be measured. These values of
$\Sigma_c$ are significantly lower (by a factor of $\gtrsim 1.5$) than the
two-sigma detection limit of \citet{ode01}. The angular radius of the smallest
distorted isodensity contour is $\sim 1^\circ$ for the orbit 6 (with the
exception of the models B1, B2, and N1), and $\gtrsim 1\fdg 8$ for other orbits.

\begin{figure*}
\plottwo{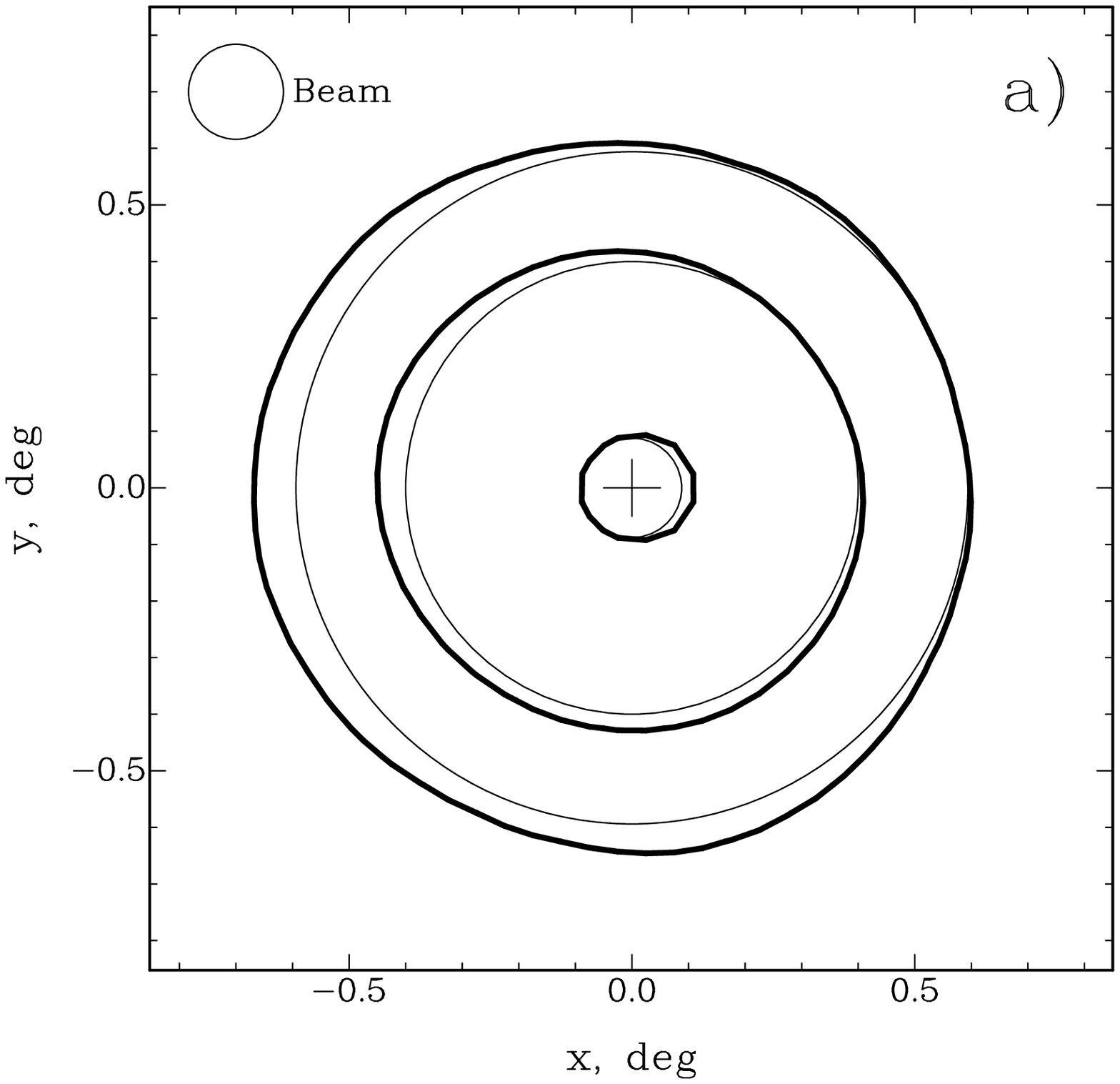}{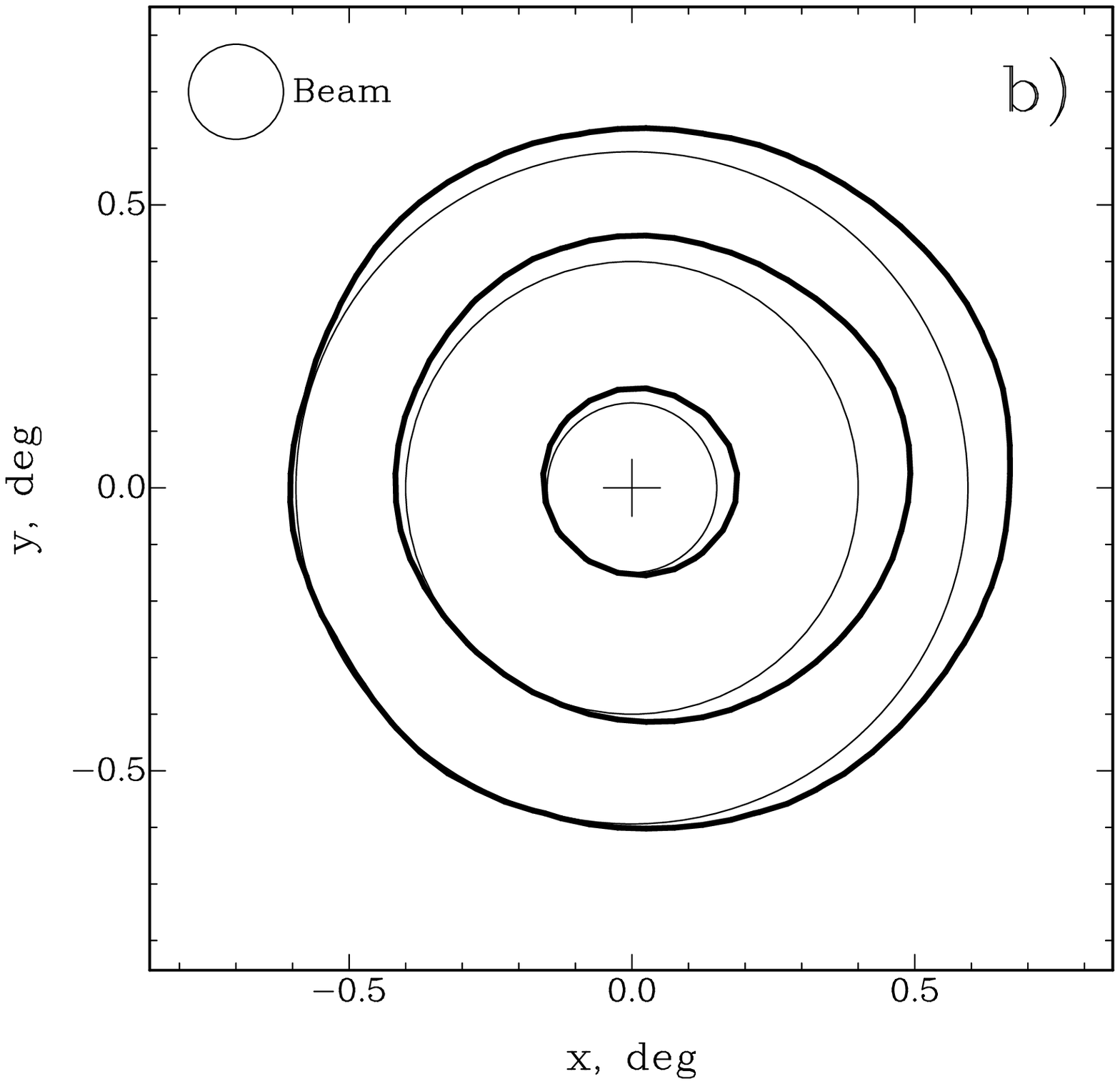}
\caption {Stellar surface brightness maps for the models B1 near the end of the
simulations (when the dwarf was $\sim 82$~kpc away from the Galactic
center). Cross marks the center of the dwarf. In the upper left corner we show
the size of the Gaussian beam used to make the maps. (a) Model B1-4. 
(b) Model B1-1.
\label{fig13} }
\end{figure*}

One very interesting special case is that of the model B1. This model has a halo
with a flat DM core of size $r_s\sim 1.4$~kpc (see Table~\ref{tab1}), so all
the observed extent of Draco is within this large harmonic core. In
Figure~\ref{fig13} we show the inner (corresponding to the observed part of
Draco) isodensity contours for the models B1-4 and B1-1. Unlike all other models
(both NFW and Burkert), here one can see a relatively strong tidal distortion of the
contours at distances $\lesssim 0\fdg5$ from the center of the dwarf.  The
distortions are substantial even for orbit 1 (Figure~\ref{fig13}b) which has
the largest pericentric distance of 70~kpc. The distorted contours are
both elliptical and non-concentric. Interestingly, the effect is minimal for the
two orbits which have the smallest eccentricity -- orbits 2 and 3 with
$R_a/R_p\leqslant 2.6$. The distortions are strong for the more eccentric
orbits 1, 4, and 5. It appears that it is the variability of the tidal force
rather than its strength which is the primary governing factor for the above effect.
As the observed stellar isodensity contours of Draco are very regular and concentric
\citep{ode01}, results of our simulations suggest that it is unlikely that
Draco resides in a large DM harmonic core -- unless it happened to move on a
close to circular orbit with $R_p\sim 45-65$~kpc and $R_a/R_p\lesssim 2.6$.

\section{DISCUSSION}

\begin{figure*}
\plottwo{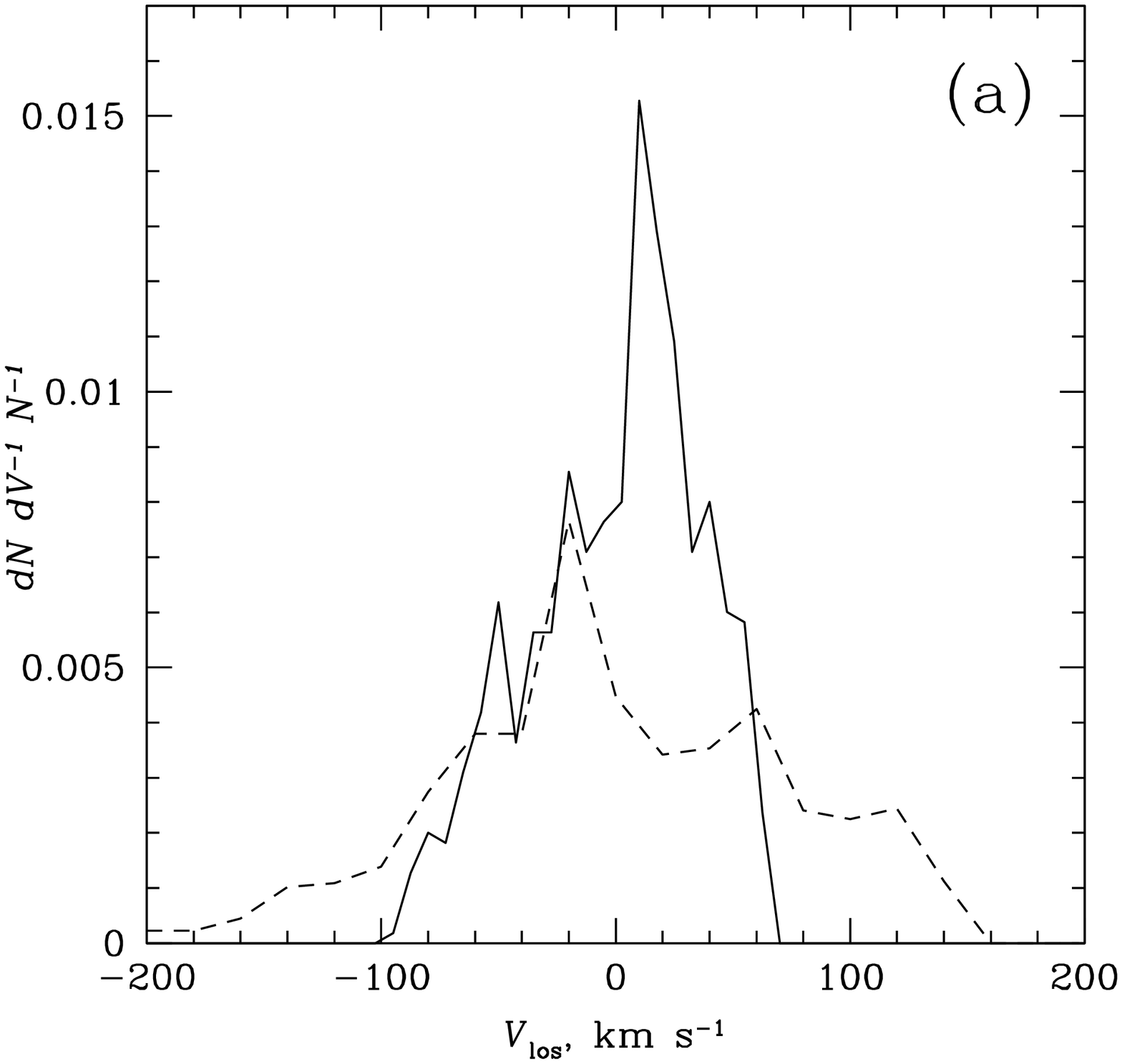}{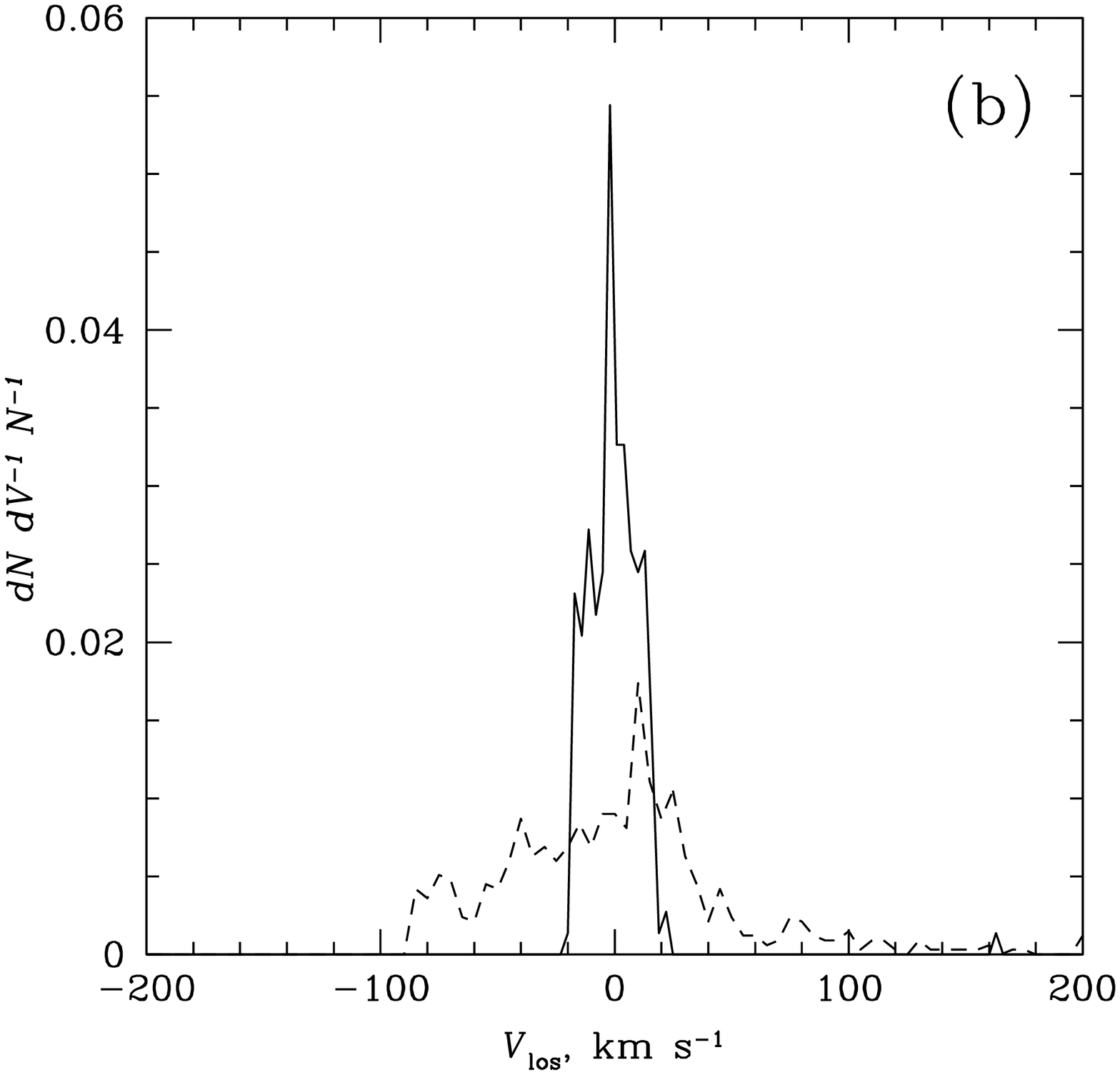}
\caption {Distribution of line-of-sight velocities $V_{\rm los}$ for the models which became
unbound by the end of the simulations. The observer is located at the center of
the Galaxy. The dwarf is located $\sim 82$~kpc away from the center of the
Galaxy. Solid and dashed lines correspond to the stars within the spatial
distance of 5 and 30~kpc from the densest part of the unbound dwarf,
respectively.  Only stars located within $0\fdg 5$ from the densest part of the
galaxy are taken into account. (a) Model B1-6.  (b) Model B2-6.
\label{fig14} }
\end{figure*}

In this paper we presented a sequence of composite (stars $+$ DM)
models for Draco, listed in Table~\ref{tab1}, which satisfy all the available
observational and cosmological constraints. We showed that for the most of
physically plausible orbits of Draco in the Galactic potential the tidal forces
could not modify the observable properties of our models appreciably after
10~Gyr of evolution. Both ``standard'' cuspy NFW DM halos and Burkert halos with
a flat core provide a reasonably good description of Draco. The properties of a
Burkert halo are better constrained by our analysis. Most importantly, if Draco
has a flat core, it should have formed at or before the end of the reionization
of the universe: $z\gtrsim 6.5$. Tidal stripping simulations put even stronger
constraints on the flat-core case: we showed in the previous section that our
most massive Burkert model B1 would be consistent with the observations only for
a very limited range of possible Draco orbits: orbit 6 is ruled out as the model
becomes completely unbound with dramatically inflated $\sigma_{\rm los}$ and
very shallow $\Sigma$ profiles, whereas for the orbits 1, 4, and 5 (and also 6)
we observe significant distortion of inner stellar isodensity contours which is
most definitely not consistent with the regular isodensity contour shape
observed in Draco by \citet{ode01}. Only the lowest eccentricity orbits 2 and
3 are not ruled out by our analysis.

An NFW halo is less constrained by the available observations of Draco: the halo
formation redshift $z$ is anywhere between $\sim 2$ and $\sim 10$, whereas the
initial virial mass could be between $\sim 10^8$ and $\sim 5\times
10^9$~M$_\odot$. In the smaller mass (and larger $z$) limit the halos are so
sturdy that even for our most disruptive orbit 6 the observable parameters of the
model can still be consistent with the Draco observations after 10~Gyr of tidal
evolution.  If Draco was accreted by the Milky Way more recently than 10~Gyr ago, the impact of
the tidal forces would be even smaller. For more massive NFW halos and for all
Burkert models the orbit 6 can be ruled out as the model predicts inflated
line-of-sight velocity dispersion in the outskirts of the dwarf which would be
at odds with observations.

How strong is our case against Draco (and other dSphs) being a ``tidal dwarf''
-- remnants of a dwarf galaxy which are not gravitationally bound at the present
time?  In the previous section we suggested that the fact that the line-of-sight
velocity dispersion is dramatically (by up to an order of magnitude) inflated in
our two ``tidal dwarf'' candidates, models B1-6 and B2-6, can be used to rule
out the ``tidal dwarf'' explanation for Draco. Here we want to caution that a
more detailed comparison between the model and observations is required to
critically assess our conclusion. Our large estimates of $\sigma_{\rm los}$ were
derived for stars with any line-of-sight velocity projected onto the dwarf disk
and optionally restricted to lie within the spatial distance of 5~kpc from the
dwarf's center. Many of the high-velocity tidal tail stars responsible for
inflating $\sigma_{\rm los}$ would be discarded by observers as ``not belonging
to the galaxy''.  In Figure~\ref{fig14} we show the distribution of
line-of-sight velocities $V_{\rm los}$ for the models B1-6 and B2-6. We show
separately histograms for freshly stripped stars (solid lines) and for all stars
projected on the disk of the dwarf (dashed lines). As one can see, the situation
depends strongly on how recently the dwarf became unbound, and on longevity of
the cold tidal streams in the Galactic halo. The model B1-6 became unbound many
orbits (almost 7~Gyr) ago, and has a very wide distribution of $V_{\rm los}$ --
even for freshly stripped stars. The model B2-6, on the other hand, became
unbound only $\sim 2$~Gyr ago, and has more complex distribution of $V_{\rm
los}$. In this model, the freshly removed stars are virtually all concentrated
within a relatively narrow interval, with $\sigma_{\rm los}$ being inflated
mainly due to the presence of one high velocity stellar particle with $V_{\rm
los}\simeq 160$~km~s$^{-1}$. Such stars will definitely be discarded by
observers. When we consider all stars (dashed line in Figure~\ref{fig14}b), the
distribution of $V_{\rm los}$ is much wider than in Galactic dSphs. In the case
of the model B2-6, the situation thus sensitively depends on how long the tidal
stream can stay collimated in the Milky Way potential. 

Another important evidence against Draco being an unbound stream of stars is
presented in Figure~\ref{fig15}. Here we show surface brightness maps for our
two unbound models B1-6 and B2-6. One can see that the contours are irregular
and not concentric -- even near the center of the dwarf. This is in sharp
contrast with the regular appearance of the observed isodensity contours in Draco
\citep{ode01}.

Pending the arrival of accurate proper motion measurements for Draco, let us be
slightly more definitive in trying to determine the nature and cosmological
significance of Draco by assuming that it is moving on the orbit 3, which is the
most probable one (see \S~\ref{Orbits}).  From Figure~\ref{fig9} one can then
infer that if Draco is a cosmological halo, its current DM mass is between
$7\times 10^7$ and $3\times 10^9$~M$_\odot$, the fraction of the tidally
stripped stars is $<3$\%, and the central line-of-sight velocity dispersion
$\sigma_0$, central surface brightness $\Sigma_0$, and the half-light radius
$r_h$ has changed due to tidal shocks by no more than $-0.07$, $-0.04$, and
$0.03$~dex, respectively, in the last 10~Gyr. This orbit has $R_p=51.1$~kpc,
placing it well outside of the Galactic disk. Stellar tidal tails produced by
our models on this orbit are extremely weak, with the surface brightness
sensitivity required to see the isodensity contours distorted due to tidal
forces being better than $\sim 200$~M$_\odot$~deg$^{-2}$, or more that two
orders of magnitude better that the observations of \citet{ode01}. The DM halo
could be either NFW or Burkert, and was formed after $z\sim 11$. Our results
then support either of the two recently proposed solutions to the ``missing
satellites'' problem \citep{kly99,moo99}: that the Galactic dSphs are the most
massive subhalos predicted by cosmological simulations to orbit in the halo of a
Milky Way sized galaxy
\citep{sto02,hay03}, or that the Galactic dSphs are the halos which managed to
form stars before the reionization of the universe was completed around $z\sim
6.5$ \citep{bul00}. Our analysis suggests that unfortunately there is not enough
of observational data yet to discriminate between the two above scenarios. Much better
quality line-of-sight velocity dispersion profiles, deeper surface brightness maps,
and ideally accurate proper motion measurements are required to produce further progress
in this direction.

We would like to mention a few most important deficiencies of our tidal
stripping model. The first one is due to our use of a spherically symmetric
potential for the Milky Way. As a result, we ignore disk shocking, which can be
very important for the orbits with $R_p\lesssim 20$~kpc. We tried to partly
circumvent this deficiency by considering an orbit with extremely small
pericentric distance: orbit 6 with $R_p\simeq 2.5$~kpc. Ideally, we would prefer
to include the disk in our simulations, but this would result in significant
increase in number of required models, which would make our project not feasible
with the current level of computing power. 

The second problem is generic to existing tidal stripping simulations of
Galactic satellites (dSphs and globular clusters), and is caused by our use of a
static potential for the Milky Way halo.  In a realistic (live) Galactic halo
very massive satellites should experience dynamical friction, which would
gradually bring them closer to the center of the Galaxy. Unfortunately, we could
not use a simple analytical formula to estimate the impact of the dynamical
friction on our results.  The dynamical friction would be strongest for the
orbits 6 and 5, which have the smallest pericentric distances.  Our subhalos on
these orbits experience dramatic mass loss (up to 90\% and 70\%, respectively)
during the first pericentric passage, rendering the constant satellite mass
formula of \citet{bin87} not applicable.  The dynamical friction equation of
\citet*{col99} does not have this limitation, but it was derived for the special
case of a subhalo with a truncated isothermal DM density profile with a core
which is very different from both the NFW and Burkert profiles of our subhalos.
We want to emphasize that an inclusion of dynamical friction in our model would
make our conclusion, that the observable properties of the most of our dwarfs
were not affected noticeable by tidal forces, even stronger, as the dwarfs would
start off at larger distance from the Milky Way center where the tidal forces
are weaker.

In addition, the use of static potential ignores the
impact of the gravitational field of the satellite on the Milky
Way halo. This effect can be very important for massive dwarfs on almost radial
orbits, with the potential of both the satellite and the Milky Way center
violently fluctuating during the pericentric passage, leading to an exchange of
energy between the dwarf and the Galactic halo (similar to the mechanism of
violent relaxation).  The above effect would probably be important only for the
orbit 6, which we were able to rule out for the most of our Draco models.

It is important to mention that our models do not cover all possible initial
configurations of Draco. In a more general case, one would have to start with
arbitrary initial stellar and DM density profiles (with the initial stellar
velocity dispersion profile following from eq.~[\ref{Jeans}]). After 10~Gyr of
evolution in the Galactic tidal field both profiles could become substantially
different, with the line-of-sight velocity dispersion either increasing (due to
the projection of unbound stars) or decreasing (due to tidal shocks; see
Fig.~\ref{fig11}). The more general case would require a dramatic increase in
supercomputing time, which would make our approach infeasible.  We want to
emphasize that despite the fact that our models probably do not include all
possible initial Draco configurations, they do constitute a family of fully
self-consistent models which match well all the available observations of
Draco.

A potentially important evolutionary factor not included in our model is an
interaction between Draco and dark subhalos, predicted to be present in the
Milky Way halo in large numbers by $\Lambda$CDM cosmological models. This effect
was studied on larger scales by \citet*{moo98}, who showed that in a cluster
environment the galaxy-galaxy harassment can be substantial. It is not clear if
the harassment on a smaller, galactic scale would be of the same order: unlike
the cluster scale, where the numbers of modeled and observed galaxies are in
good agreement, there is a ``missing satellites'' issue on galactic scales.

\section{CONCLUSIONS}

We presented two one-parameter families (separately for NFW and Burkert DM
density profiles) of composite (stars $+$ DM) models for Draco which satisfy all
the available observational and cosmological constraints. We showed that these
models can survive tidal shocks and stripping on most realistic Draco orbits in
the Galactic potential for 10~Gyr, with no appreciable impact on their
observable properties.  Both NFW and Burkert DM halo profiles are found to be
equally plausible for Draco. The DM halos are either massive (up to $\sim
5\times 10^9$~M$_\odot$) and recently formed ($z\sim 2\dots 7$), or less massive
(down to $\sim 7\times 10^7$~M$_\odot$) and older ($z\sim 7\dots 11$).
Consequently, our results can be used to support either of the two popular
solutions of the missing satellites problem -- ``very massive dwarfs'' and
``very old dwarfs''.  Higher quality observations (line-of-sight velocity
dispersion profiles, surface brightness maps, proper motion measurements) are
required to further constraint the properties of Galactic dSphs and to place
them in the right cosmological context.

\acknowledgements 
We would like to thank Jan Kleyna for providing the observed line-of-sight velocity
dispersion profile for Draco.  The simulations reported in this paper were carried
out on McKenzie cluster at the Canadian Institute for Theoretical Astrophysics.

\appendix
\section{Derivation of space velocity vector components for Galactic satellites}
\label{prop}

In this section we derive the components of the space velocity vector of an
object with known distance from the Sun $D$, heliocentric line-of-sight velocity
$V_{\rm los}$, and two proper motion components $\mu_\alpha \cos\delta$ and
$\mu_\delta$. We work with the frame of reference where the center is at the
Sun, the axis $X$ is directed toward the Galactic center, the axis $Y$ is
pointing at ($l=90\degr$, $b=0$), and the axis $Z$ is directed toward the North
Galactic Pole ($b=90\degr$). Here ($l,b$) are the galactic coordinates. The
frame of reference is at rest relative to the Galactic center.  The Solar
velocity vector in this frame of reference is ${\bf
V_\odot}=\{10.0,225.25,7.17\}$~km~s$^{-1}$
\citep{deh98}. We assume that the Sun is located at the distance $R_\odot=8.5$~kpc 
from the Galactic center.

To convert the proper motion vector components from equatorial to galactic
frame of reference, one can use

\begin{eqnarray}
\mu_l \cos b &=& (\mu_\alpha \cos\delta)\cos\varphi - \mu_\delta \sin\varphi,\\
\mu_b &=& (\mu_\alpha \cos\delta)\sin\varphi + \mu_\delta \cos\varphi,
\end{eqnarray}

\noindent where the angle $\varphi$ is obtained from

\begin{eqnarray}
\cos\varphi &=& (\sin\delta_{\rm NGP}-\sin\delta\sin b)/(\cos\delta \cos b), \\
\sin\varphi &=& -\sin(\alpha-\alpha_{\rm NGP})\cos\delta_{\rm NGP}/\cos b.
\end{eqnarray}

\noindent Here ($\alpha,\delta$) and ($l,b$) are the equatorial and galactic coordinates
of the object, respectively, and ($\alpha_{\rm NGP},\delta_{\rm NGP}$) are
the equatorial coordinates of the North Galactic Pole (for the $J2000$ equinox,
$\alpha_{\rm NGP}=192\fdg 859$ and $\delta_{\rm NGP}=27\fdg 128$).

One can show that the three components of the space velocity vector of the object
in the frame of reference moving with the Sun are

\begin{eqnarray}
U_z &=& zV_{\rm los} + D\mu_b(x^2+y^2)^{1/2},\\
U_x &=& [x(V_{\rm los}-zU_z) - D(\mu_l\cos b)y(x^2+y^2)^{1/2}]/(x^2+y^2),\\
U_y &=& [D(\mu_l\cos b)(x^2+y^2)^{1/2} + yU_x]/x,
\end{eqnarray}

\noindent where $x=\cos l \cos b$, $y=\sin l \cos b$, and $ z=\sin b$ are the
components of a unit vector directed from the Sun toward the object, and the
units for $D$, $\bf U$, and the two proper motion components ($\mu_l\cos
b,\mu_b$) are km, km~s$^{-1}$, and rad~s$^{-1}$, respectively.

In the frame of reference which is at rest relative to the Galactic center, the
space velocity vector of the object is ${\bf V}={\bf U} + {\bf V_\odot}$.  In
the cylindrical Galactic frame of reference, the three components of the space
velocity vector of the object are

\begin{equation}
\Pi=(S_xV_x+S_yV_y)/(S_x^2+S_y^2)^{1/2}, \quad \Theta=(S_yV_x-S_xV_y)/(S_x^2+S_y^2)^{1/2},
\quad W=V_z,
\end{equation}

\noindent where $\Pi$ is directed outward from the Galactic center in the plane of the Galaxy,
$\Theta$ is the circular rotation speed in the plane of the Galaxy (positive for
the Sun), $W$ is directed toward the North Galactic Pole, and ${\bf
S}=\{xD-R_\odot,yD,zD\}$ is the vector connecting the Galactic center with the
object. In the spherical Galactic frame of reference, the radial and tangential
velocities of the object are

\begin{equation}
V_r=({\bf V}\cdot {\bf S})/|{\bf S}|, \quad V_t=(|{\bf V}|^2-V_r^2)^{1/2}.
\end{equation}

\end{document}